\documentclass[aps,prd,twocolumn,superscriptaddress,nofootinbib,preprintnumbers]{revtex4-1}

\usepackage{graphicx,xcolor,amsmath,amssymb}
\usepackage{slashed}

\usepackage{hyperref} 
\usepackage{amsmath}
\hypersetup{
    colorlinks=true,
    citecolor=[rgb]{.1, .7, .6},
    linkcolor=[rgb]{.2, .55, .95},
    filecolor=magenta,
    urlcolor=[rgb]{.1, .7, .6},
}

\newcommand{\Tab}[1]{Table~\ref{#1}}
\newcommand{\ie}{{i.e.~}}  
\newcommand{\eg}{{e.g.~}}
\newcommand{\Sec}[1]{Sec.~\ref{#1}}
\newcommand{\Fig}[1]{Fig.~\ref{#1}}
\newcommand{\Eq}[1]{Eq.~\eqref{#1}}

\definecolor{mypurple}{RGB}{164,64,214}

\begin{document}

\title{Transient Radio Lines from Axion Miniclusters and Axion Stars }

\newcommand{\GRAPPA}{\affiliation{
    GRAPPA Institute, 
    Institute for Theoretical Physics Amsterdam and Delta Institute for Theoretical Physics,
    University of Amsterdam, 
    Science Park 904, 
    1098 XH Amsterdam, 
    The Netherlands
    }}

\newcommand{\OKC}{\affiliation{
    The Oskar Klein Centre, 
    Department of Physics, 
    Stockholm University, 
    AlbaNova, 
    SE-10691 Stockholm, 
    Sweden
    }}

\newcommand{\NORDITA}{\affiliation{
    Nordita, 
    KTH Royal Institute of Technology and Stockholm University, 
    Roslagstullsbacken 23, 
    10691 Stockholm, 
    Sweden
    }}

\newcommand{\SITP}{\affiliation{
    Stanford Institute for Theoretical Physics, 
    Department of Physics, 
    Stanford University, 
    Stanford, CA 94305, 
    USA
    }}

\author{Samuel J. Witte}
\email{s.j.witte@uva.nl}
\GRAPPA

\author{Sebastian Baum}
\email{sbaum@stanford.edu}
\SITP

\author{Matthew Lawson}

\author{M.C.~David Marsh}
\OKC

\author{Alexander J. Millar}\email[]{amillar@fnal.gov}
\OKC
\NORDITA
  \affiliation{Fermi National Accelerator Laboratory, Batavia, Illinois 60510, USA}

\author{Gustavo Salinas}
\email{gustavo.salinas@fysik.su.se}
\OKC

\begin{abstract}
 Gravitationally bound clumps of dark matter axions in the form of `miniclusters' or even denser `axion stars' can generate strong radio signals through axion-photon conversion when encountering highly magnetised neutron star magnetospheres. We systematically study encounters of axion clumps with neutron stars and characterise the axion infall, conversion and the subsequent propagation of the photons. We show that the high density and low escape velocity of the axion clumps lead to strong, narrow, and temporally characteristic  transient radio lines with an expected duration varying from seconds to months. Our work comprises the first end-to-end modeling pipeline capable of characterizing the radio signal generated during these transient encounters, quantifying the typical brightness, anisotropy, spectral width, and temporal evolution of the radio flux. The methods developed here may prove essential in developing dedicated radio searches for transient radio lines arising from miniclusters and axion stars. 
\end{abstract}

\maketitle
\preprint{FERMILAB-PUB-22-908-T}

\section{Introduction}

Despite comprising over $25\%$ of the energy density in the Universe, the fundamental nature of dark matter 
remains unknown~\cite{collaboration2020planck}. Among the most well-motivated candidates for dark matter is the axion -- a light pseudoscalar arising from a broken global $U(1)$ symmetry that was originally introduced to solve the strong CP problem (\ie the question of why quantum chromodynamics seems to conserve charge-parity symmetry)~\cite{Peccei:1977hh,Peccei:1977ur,Weinberg:1977ma,Wilczek:1977pj}.

Axion dark matter can be abundantly produced via various non-thermal processes in the early Universe, including the misalignment mechanism and the decays of topological defects (\ie cosmic strings and domain walls)~\cite{Preskill:1982cy,Abbott1982,Fischler1982,Davis:1986xc,Lyth:1991bb}. In the event that the global $U(1)$ Peccei-Quinn symmetry is broken after the end of inflation, one expects these production mechanisms to generate modest $\mathcal{O}(1)$ fluctuations in the axion density field; the produced over-densities can subsequently undergo gravitational collapse near matter-radiation equality to source small virialized structures known as `axion miniclusters'~\cite{Hogan:1988mp,Kolb:1993zz,Kolb:1993hw,Kolb:1994fi,Kolb:1995bu,Zurek:2006sy,Hardy:2016mns,OHare:2017yze,Dokuchaev:2017,Fairbairn:2017sil,Vaquero:2018tib,Eggemeier:2019khm,Buschmann:2019icd,Xiao:2021nkb,Ellis:2022grh}. As the miniclusters subsequently relax, the central regions of these objects can condense and form stable high-density cores known as `axion stars'~\cite{Kaup:1968zz,Ruffini:1969qy,Tkachev:1986tr,Kolb:1993zz,Seidel:1993zk,Barranco:2010ib, Levkov:2018kau,Eggemeier:2019jsu,Chen:2020cef,Braaten:2015eeu, Schiappacasse:2017ham, Visinelli:2017ooc}.

Recent years have shown significant progress in simulating the dynamics of the axion field in the early Universe, leading to estimates of the fraction of axion dark matter that collapses into miniclusters $f_{\rm AMC} \sim \mathcal{O}(1-100)\,\%$~\cite{Vaquero:2018tib,Buschmann:2019icd,Xiao:2021nkb} in the post-inflationary Peccei-Quinn symmetry breaking scenario. A majority of these miniclusters are expected to survive until today -- the notable exception being those which pass through dense stellar environments (such as found in the center of galaxies), where stellar encounters can efficiently strip and disrupt even the densest cores ~\cite{Kavanagh:2020gcy,Dandoy:2022prp,Shen:2022ltx}. Notice that the value of $f_{\rm AMC}$ plays an important role in determining how and where to search for axion dark matter;  in the limit $f_{\rm AMC} \rightarrow 100\%$, the local dark matter density is expected to tend toward zero (this is a consequence of the fact that Earth is statistically unlikely to be embedded in such an object), meaning the sensitivity of laboratory-based searches for axion dark matter may be severely reduced, or even eliminated entirely (although a recent study has suggested that the local dark matter density may only be reduced by $\mathcal{O}(90\%)$~\cite{Eggemeier:2022hqa}). In such a scenario, one may have to rely entirely on indirect searches, and in particular those which account for the stochastic nature of the underlying dark matter distribution.

\begin{figure*}
    \centering
    \includegraphics[width=0.49\textwidth]{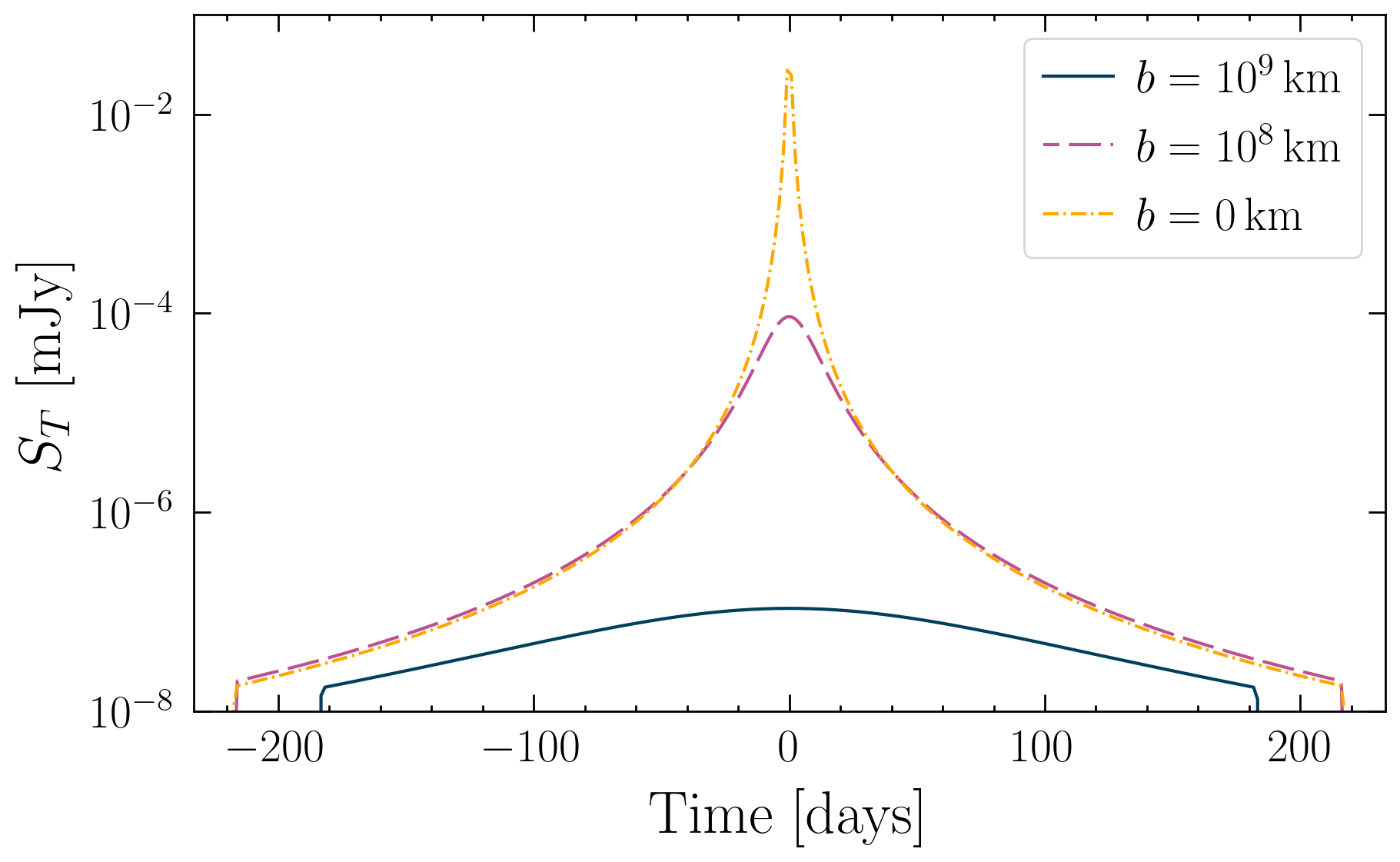}
    \includegraphics[width=0.49\textwidth]{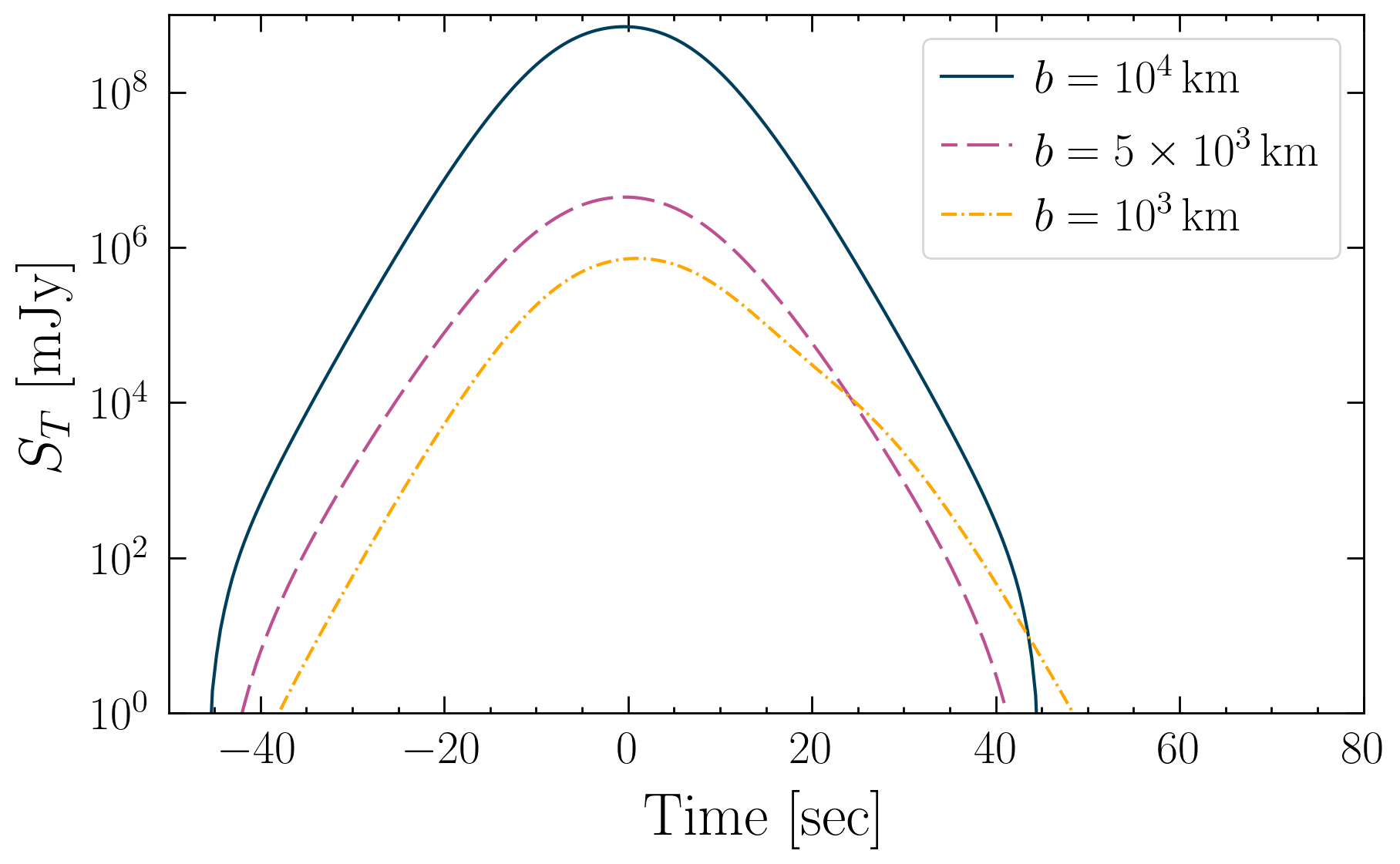}
    \caption{Long term time evolution of the sky-averaged differential flux for the fiducial minicluster-neutron star (left) and axion star-neutron star (right) encounter at various values of the impact parameter $b$. All other parameters describing the axion minicluster/star, the neutron star, and the encounter are fixed to the fiducial values in \Tab{tab:fid}. For the minicluster encounter, we expect the maximum transient time to scale $t_{\rm max} \propto M_{\rm AMC}^{1/3}$, while for the axion star we expect it to scale as $t_{\rm max} \propto M_{\rm AS}^{-1}$ (for a fixed axion mass). }
    \label{fig:long_time}
\end{figure*}

One of the more promising proposals to try and indirectly search for the existence of axions is to look for radio signatures produced from axion-photon mixing in the magnetospheres of neutron stars~\cite{Pshirkov:2007st,Huang:2018lxq,Hook:2018iia,Safdi:2018oeu,Battye:2019aco,Leroy:2019ghm,Foster:2020pgt,Witte:2021arp,Battye:2021xvt,Millar:2021gzs,Foster:2022fxn,Noordhuis:2022ljw}. Owing to the strong magnetic fields, the mixing in these environments is large, and the presence of a spatially varying background plasma in neutron star magnetospheres can enable {\it resonant} conversion, occurring when the axion mass approximately matches the effective mass of photons in the plasma~\cite{Raffelt:1987im}. In some cases, this resonance can even lead to $\mathcal{O}(1)$ axion-to-photon conversion probabilities (see \eg~\cite{Foster:2022fxn}). This field has seen significant progress over the last few years, with major improvements on \eg the computation of axion-photon mixing in highly magnetized plasma~\cite{Millar:2021gzs}, and the impact of refractive, dispersive, and dissipative effects of the plasma on the expected radio signal~\cite{Witte:2021arp,Battye:2021xvt}.

A number of dedicated radio searches for axion dark matter have already been performed; these searches have used various telescopes and targeted both individual neutron stars and the broader neutron star population in the Galactic Center\footnote{It is worth noting that radio observations of pulsars can also be used to constrain axions even if they do not contribute to the dark matter~\cite{Prabhu:2021zve,Noordhuis:2022ljw}.}, leading to competitive constraints on the axion-photon coupling across a range of axion masses~\cite{Foster:2020pgt,Battye:2021yue,Foster:2022fxn}\footnote{As a word of caution, we note the constraints derived from these searches cannot be directly compared, as the implicit assumptions entering the modeling yield significant changes to the inferred limits (see \eg comparisons made in~\cite{Foster:2022fxn}).}. An implicit assumption in these searches is that axions are smoothly distributed in the inner parts of the galaxy, meaning the spectral line is approximately static when averaged on timescales much longer than the rotational period of the pulsar. Should, however, the number density of axion clumps (henceforth, we will use the term axion clump to interchangeably refer to both miniclusters as axion stars) be non-negligible, one instead expects the appearance of transient radio lines, which are generated as these objects pass through the neutron star magnetospheres~\cite{Edwards:2020afl,iwazaki2015axion,Buckley:2020fmh,Nurmi:2021xds,Bai:2017feq,dietrich2019neutron,Bai:2021nrs,Kouvaris:2022guf}. The  expectation is that the large dark matter densities found in these gravitationally bound objects will allow one to probe small values of the axion-photon coupling, potentially even testing the parameter space of the QCD axion (see \eg~\cite{Edwards:2020afl}). This scenario, however, is far more difficult to treat than the case of the smooth axion background, as the observability of these transients depend 
on: $(i)$ the properties and distributions of axion clumps at formation, $(ii)$ the tidal stripping and disruption of these objects at late times, $(iii)$ the properties and distributions of the neutron star population, and $(iv)$ the non-linear dynamics of each individual encounter (from the tidal disruption and in-fall, to the photon production and propagation). The focus of this work is on developing the tools and formalism required to treat $(iv)$, providing a crucial step toward understanding how to develop and optimize future radio searches for miniclusters and axion stars.

Before delving into details, let us first provide a high-level overview of the properties and characteristics of these transient events. The radio emission from an interaction of an axion clump with a neutron star can endure on timescales spanning from seconds to years, depending on the characteristic size of the axion clump. During the encounter, one expects a steady rise in the flux, followed by an extended fall (see \eg \Fig{fig:long_time} for illustrative examples\footnote{Notice that the right panel of \Fig{fig:long_time} illustrates a rather unintuitive scaling of flux with the impact parameter, with larger values of $b$ generating stronger signals -- this effect arises as a consequence of the the internal velocity dispersion in the axion star, and is discussed in \Sec{sec:results}.}), with the temporal evolution set by the density profile and the impact parameter. Notice that for typical velocities $v\sim\mathcal{O}(100)$ km/s, the maximum impact parameter leading to radio emission is roughly comparable to the size of the axion clump itself; for the fiducial models shown in  \Fig{fig:long_time}, this maximum impact parameter roughly corresponds to the blue line. In addition to the long term time evolution of the signal, one expects time structure to appear at the level of the rotational period of the neutron star (spanning from $\sim 0.1\textit{--}10$ seconds for a typical pulsar, which is typically much less than the transient timescale); this is a consequence of having $(i)$ a misalignment between the magnetic and rotational axes, and $(ii)$ a highly asymmetric axion phase space near the neutron star. The asymmetry in the axion phase space is also expected to generate a highly inhomogeneous radio signal, especially in the case of small clumps, which may only illuminate a small fraction of the sky. Finally, it is worth highlighting that the central densities of these objects can be extremely large, meaning that they may be observable to extragalactic, or even cosmological, distances (depending on the details of the object and the axion-photon coupling). For example, consider the peak flux density produced from the axion star-neutron star encounter shown in \Fig{fig:long_time}: for an impact parameter of $b = 10^4\,$km at a distance of 1 kpc from Earth (and fixing parameters to the fiducial values shown in \Tab{tab:fid}), we find a peak flux of $S \sim 10^9\,$mJy; assuming a telescope sensitivity of $\mathcal{O}(10)$ mJy (roughly consistent with the minimum flux density observed from a fast radio burst, see ~\cite{petroff2016frbcat}), this event could  be observed out to distances of $\sim 10$ Mpc, \ie anywhere in the Local Group.

The organization of this paper is as follows. In \Sec{sec:mini_stars} we describe the current knowledge of axion miniclusters and axion stars -- including their formation, properties, and evolution over the cosmic history. In \Sec{sec:grav_ph} we outline the formalism used to treat the gravitational disruption of these objects, as well as photon production and propagation. Finally, we present the results of this analysis in \Sec{sec:results}, illustrating the characteristic strength of the radio signal, its anisotropy on the sky, the time-domain structure, the spectral properties, and the sensitivity of these quantities to \eg the impact parameter and relative velocities. We conclude in \Sec{sec:conc}.

\section{Axion Miniclusters and Axion Stars} \label{sec:mini_stars}

In this section, we outline the current knowledge of axion miniclusters and axion stars, motivating in particular the characteristic properties used to compute the radio signals generated in the remainder of this paper.

\subsection{Axion Miniclusters}
An axion minicluster is a self-gravitating virialized clump of axions~\cite{Hogan:1988mp,Kolb:1993zz,Kolb:1994fi,Kolb:1995bu, Zurek:2006sy}. Although axion miniclusters have been discussed for more than four decades, many of their properties are still unclear. They form from inhomogeneities in the initial axion density field and thus are mostly relevant in the {\it post-inflationary} Peccei-Quinn (PQ) symmetry breaking scenario. Numerical simulations of the axion field through the QCD phase transition~\cite{Kolb:1993hw,Vaquero:2018tib,Buschmann:2019icd} as well as from matter-radiation equality until redshift $z \sim 100$~\cite{Eggemeier:2019khm} have been carried out to infer the mass function as well as the density profiles of axion miniclusters. While we have learned much about miniclusters in the post-inflationary PQ scenario from these simulations, there are still many open questions. Among the properties most relevant for this work are the minicluster density profiles,  which are expected to sit somewhere between a power law and a broken power law. The formation of the most massive axion miniclusters arguably involves hierarchical merging; in this case one would expect their density profile to be described by a Navarro-Frenk-White (NFW) profile~\cite{Navarro:1995iw}
\begin{equation} \label{eq:dens_NFW}
   \rho(r) = \frac{\rho_s}{\frac{r}{r_s} \left( 1 + \frac{r}{r_s} \right)^2} \;,
\end{equation}
where $\rho_s$ is the characteristic density and $r_s$ the scale radius. On the other hand, the initial formation mechanism of axion miniclusters from the overdensities imprinted in the axion field is more akin to a direct collapse when gravity starts to be relevant around matter-radiation equality~\cite{Zurek:2006sy, OHare:2017yze}. In this case, the density profile should be given by a power-law~\cite{Bertschinger:1985pd,Zurek:2006sy}
\begin{equation} \label{eq:dens_PL}
   \rho(r) = \rho_s \left( \frac{r_s}{r} \right)^{9/4}\;.
\end{equation} 

Assuming that the miniclusters formed from direct collapse shortly before matter radiation equality, their characteristic density is given in terms of the overdensity $\delta$~\cite{Kolb:1994fi}
\begin{equation}
   \rho_s = 140 \left( 1 + \delta \right) \delta^3 \rho_{\rm eq} \;,
\end{equation}
where $\rho_{\rm eq}$ is the averaged density of axions at matter-radiation equality. Should axions comprise the entirety of the dark matter, $\rho_{\rm eq} = 1.3 \times 10^3 \,M_\odot/{\rm pc}^3$ (assuming Planck cosmology~\cite{Planck:2018vyg}). From the simulations of the axion field through the QCD phase transition, one obtains a typical values of $\delta \sim 1$~\cite{Buschmann:2019icd} at matter-radiation equality. Hence, a typical value for the density scale of axion miniclusters is $\rho_{\rm AMC} \sim 10^5 M_\odot/{\rm pc}^3$, although $\rho_{\rm AMC}$ is obviously very sensitive to the value of $\delta$. The density profiles of axion miniclusters in the Milky Way will also be affected by encounters with stars~\cite{Dokuchaev:2017,Kavanagh:2020gcy,Dandoy:2022prp, Shen:2022ltx}, although we neglect such effects here.

In order to relate the mass of the minicluster $M_{\rm AMC}$ and its typical density $\rho_{\rm AMC}$ to those of a NFW or power-law density profile we follow Refs.~\cite{Fairbairn:2017sil,Kavanagh:2020gcy}. For the NFW profile, we identify
\begin{equation}
   \rho_s^{\rm NFW} = \rho_{\rm AMC} \;,\quad r_s^{\rm NFW} = \left( \frac{M_{\rm AMC}}{4\pi \rho_{\rm AMC} f(c)} \right)^{1/3} \;,
\end{equation}
where $f(c) = \ln(1+c) - c/(1+c)$ and we choose a concentration parameter $c=100$. We truncate the NFW profile at $R_{\rm AMC}^{\rm NFW} = c r_s$. For the power-law profile, we choose a truncation radius
\begin{equation}
   R_{\rm AMC}^{\rm PL} = \left( \frac{3 M_{\rm AMC}}{4\pi \rho_{\rm AMC}} \right)^{1/3} \;,
\end{equation}
and fix the normalization of the profile, \Eq{eq:dens_PL} via $\rho_s^{\rm PL} \left(r_s^{\rm PL}\right)^{9/4} = \frac{1}{4} \rho_{\rm AMC} \left( R_{\rm AMC}^{\rm PL} \right)^{9/4}$. 

For the velocity profile of axions in the minicluster, we assume a Maxwell-Boltzmann distribution with velocity dispersion $\sigma_v(r)$ truncated at the escape velocity $v_{\rm esc}$, 
\begin{equation} \label{eq:f(v)MB}
   \tilde{f}(\vec{v}) = \frac{1}{N_{\rm esc}} \left( \frac{1}{2\pi \sigma_v^2} \right)^{3/2} e^{-\frac{v^2}{2\sigma_v^2}} ~H\left(v-v_{\rm esc}\right)\,,
\end{equation}
in the minicluster rest frame. In \Eq{eq:f(v)MB}, $H(x)$ denotes the Heaviside step function. The normalization coefficient is given by
\begin{equation}
   N_{\rm esc} = {\rm erf} \left( \frac{v_{\rm esc}}{\sqrt{2} \sigma_v} \right) - \sqrt{\frac{2}{\pi}} \frac{v_{\rm esc}}{\sigma_v} e^{- \frac{ v_{\rm esc}^2}{2\sigma_v^2}} \;.
\end{equation} 
For a virialized minicluster with spherically symmetrical density distribution $\rho(r)$, the velocity dispersion is given by the circular velocity, $\sigma_v(r) = \sqrt{G M(r) / r}$ where~\cite{BinneyTremaine:1987} 
\begin{equation}
    M(r) = 4 \pi \int_0^r dr'\; r'^2 \rho(r') \;,
\end{equation}
is the mass enclosed within the radius $r$. Similarly, the escape velocity $v_{\rm esc}(r) = \sqrt{ \left| 2 \Phi (r) \right|}$ is given by the gravitational potential at $r$,
\begin{equation}
    \Phi(r) = - 4\pi G \left[ \frac{1}{r} \int_0^r dr' \; r'^2 \rho(r') + \int_r^\infty dr'\; r' \rho(r') \right]\;.
\end{equation}

For the numerical results shown in this work, we consider axion miniclusters with an NFW profile. Furthermore, we neglect the radial dependence of the escape velocity and velocity dispersion, and set these quantities to their respective values evaluated at $R_{\rm AMC}$. This is done in order to simplify the sampling procedure, however it is worth emphasizing that this can be corrected in a straight-forward manner by employing an importance sampling scheme. We have validated that for the examples shown this approximation has a negligible impact on the radio flux. 

\subsection{Axion Star}
Axion stars~\cite{Kaup:1968zz,Ruffini:1969qy,Tkachev:1986tr} are, like axion miniclusters, self-gravitating clumps of axions. However, while axion miniclusters are virialized objects formed from the direct collapse of overdensities in the axion field or hierarchical mergers of such direct-collapse miniclusters, axion stars are equilibrium solutions of the classical axion field equations found by balancing the self-gravity of the (classical) field describing an axion configuration with the gradient pressure of that field. It has been shown in numerical simulations that axion stars form efficiently in the centers of axion miniclusters~\cite{Levkov:2018kau,Eggemeier:2019jsu,Chen:2020cef}, although in what follows, for simplicity, we will treat axion stars as isolated objects. 

A heuristic explanation of the properties of axions stars in the classical field description based on balancing self-gravity with the gradient pressure of the field can, e.g., be found in Ref.~\cite{Visinelli:2017ooc}. One can also understand the properties of an axion star heuristically in the axion particle picture: Let us denote the mass and radius of the axion star configuration as $M_{\rm AS}$ and $R_{\rm AS}$, respectively, and the typical velocities of axions in the axion star as $v$. The contributions of the axions' kinetic energy and the self-gravity and to the axion star's energy are then
\begin{equation} \label{eq:UAS}
    U_{\rm AS} \sim M_{\rm AS} v^2 - \frac{G M_{\rm AS}^2}{R_{\rm AS}} \;.
\end{equation}
Since the axions are confined in a volume with linear dimensions $R_{\rm AS}$, their velocities must at least be as large as what is dictated by the uncertainty principle, $v \sim \hbar / m_a R_{\rm AS}$, where $m_a$ is the axion mass.\footnote{Equivalently, one can note that the size of an axion configuration with velocity $v$ must be at least as large as the axions' De Broglie wavelength, $R_{\rm AS} \gtrsim \hbar / m_a v$.} If one substitutes this relation into \Eq{eq:UAS}, one finds that the energy of the configuration is minimized for
\begin{equation}
   R^{\rm AS}_{90} = \frac{\alpha_k \hbar^2}{G m_a^2} \frac{1}{M_{\rm AS}} \;.
\end{equation}
In order to fit this heuristic result to numerical solutions of axions, we have here replaced $R_{\rm AS}$ with $R^{\rm AS}_{90}$, the radius containing 90\,\% of the axion star's mass, and inserted a numerical coefficient, $\alpha_k \sim 10$~\cite{Visinelli:2017ooc}. 

Beyond the characteristic mass-radius relations of axion stars, $R_{\rm AS} \propto 1/{M_{\rm AS}}$, we can learn two more lessons about axion stars from this heuristic argument: First, the gradient pressure stabilizing axion stars in the classic field picture has its origin in the particle nature of axions and the uncertainty principle; note that in the literature this gradient pressure is sometimes referred to as ``quantum pressure''. Second, the heuristic argument suggests that axion stars are the densest axion configurations possible held together by gravity. While denser axion configurations are possible if they are bound by forces other than gravity, for example, the characteristic attractive quartic self-interactions of axions, such denser configurations are unstable to perturbations and/or decay quickly~\cite{Schiappacasse:2017ham,Visinelli:2017ooc}. The instabilites brought about by the self-interactions also set a maximal mass for (stable) axions stars,
\begin{equation} \label{eq:AS_maxm}
   M_{\rm AS}^{\rm max} \simeq 7.0 \times 10^{-12}\,M_\odot \left( \frac{f_a}{6 \times 10^{11}\,{\rm GeV}} \right) \left(\frac{10\,\mu{\rm eV}}{m_a}\right) \;,
\end{equation}
where $f_a$ is the axion decay constant controlling the self-interactions.

\begin{figure*}
    \centering
    \includegraphics[width=0.48\textwidth]{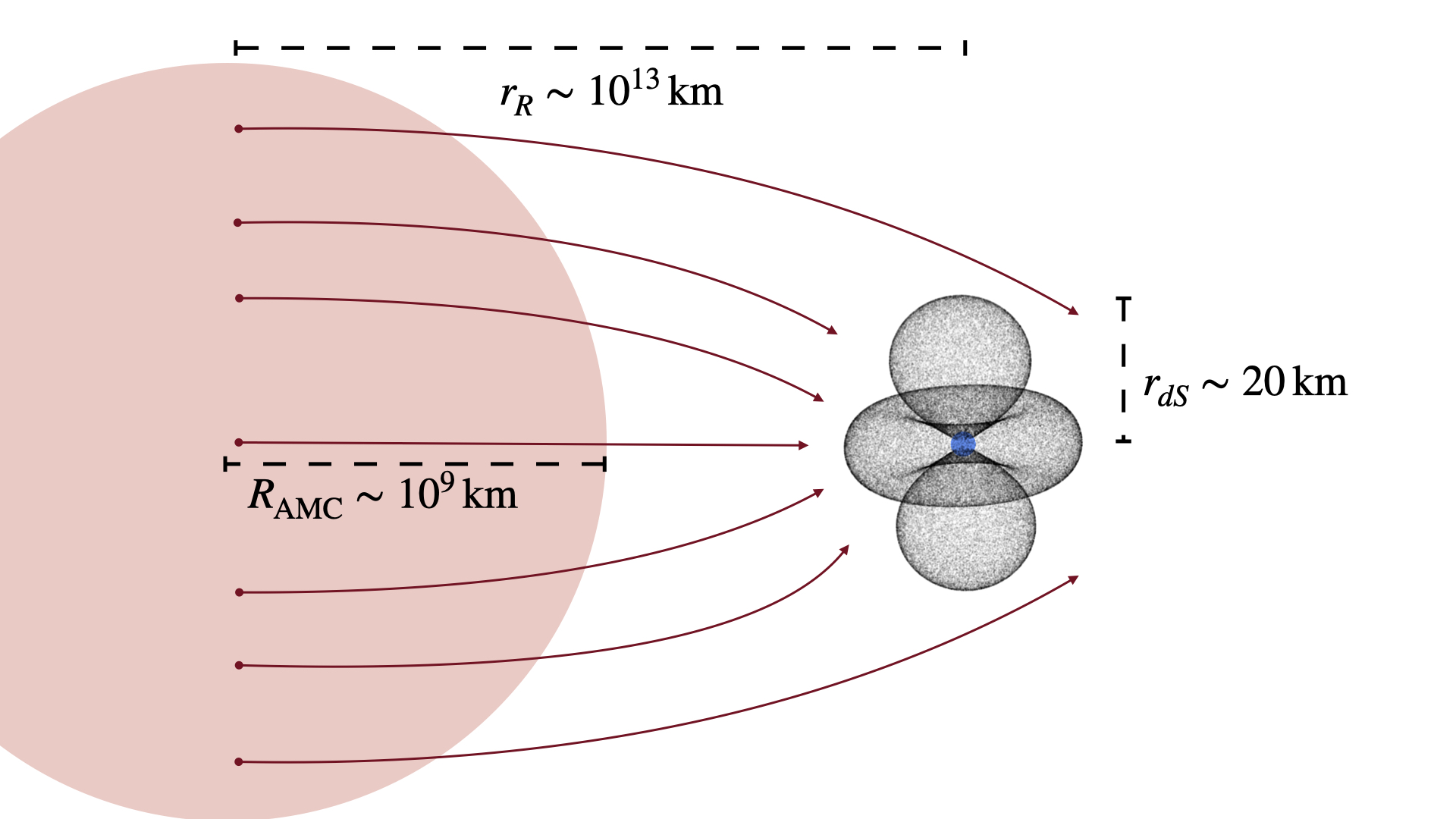}
    \includegraphics[width=0.48\textwidth, trim={0cm 0cm 0cm 0cm}, clip]{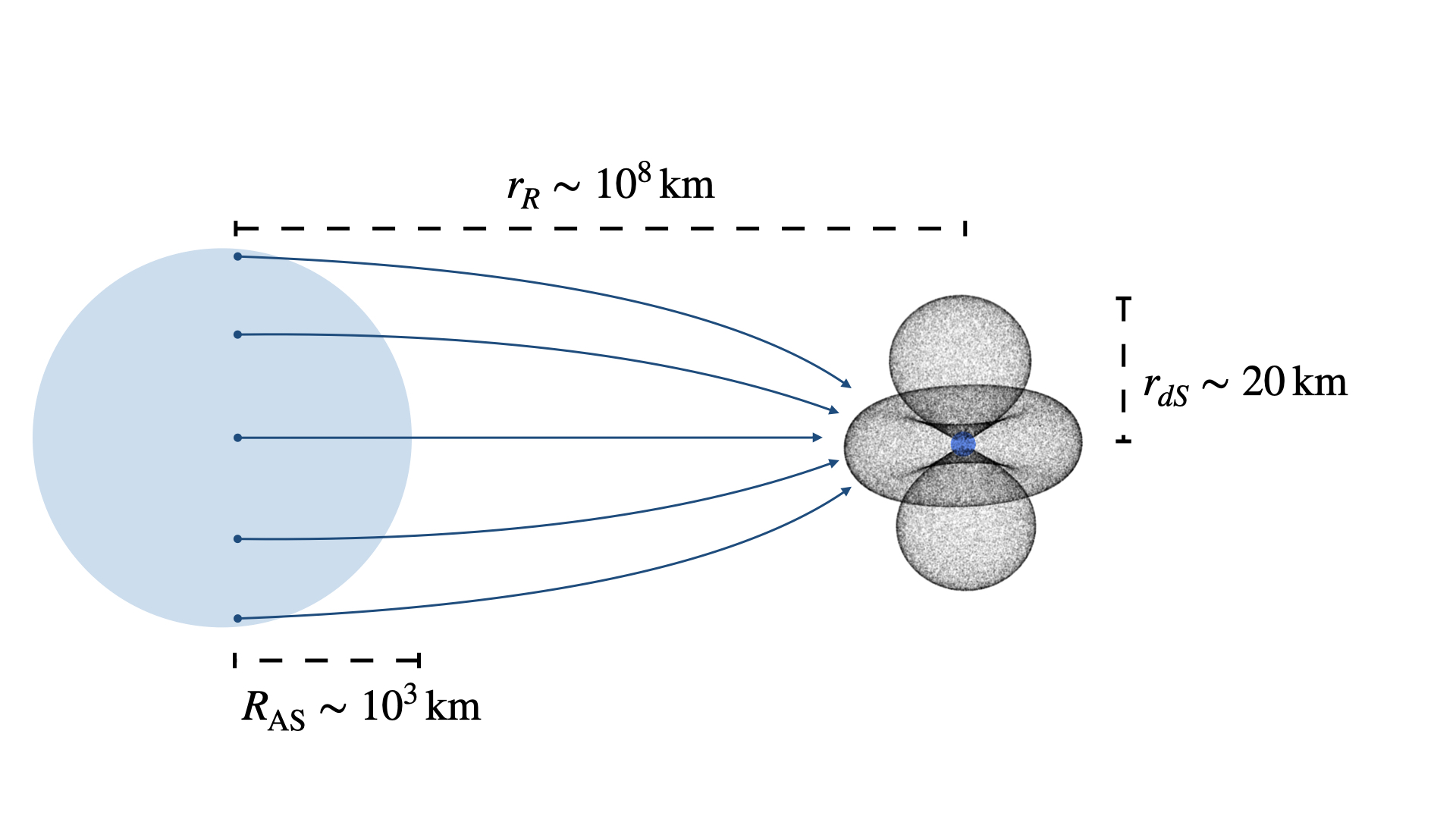}
    \caption{Illustration of spaghettified axion phase space arising during minicluster- (left) and axion star-neutron star (right) encounter. In each case, we highlight three characteristic scales (as computed under the fiducial models): the Roche radius $r_R$, the radius of the minicluster or axion star $R_{\rm AMC/AS}$, and the characteristic scale of the conversion surface $r_{dS}$. In both cases the conversion surface is shown in black, surrounding the neutron star (small blue dot at center). }
    \label{fig:spag}
\end{figure*}

The precise density profile of axion stars can only be determined from numerical solutions to the axion equations of motion; in general, a ``sech'' Ansatz has been shown to yield a good fit to the numerical solutions over a wide range of masses~\cite{Schiappacasse:2017ham},
\begin{equation} \label{eq:dens_sech}
  \rho(r) = \frac{3}{\pi^3} \frac{M_{\rm AS}}{(R^{\rm AS}_{\rm sech})^3} \frac{1}{\cosh^2(r/R^{\rm AS}_{\rm sech})} \;,
\end{equation} 
where $R^{\rm AS}_{\rm sech} = 0.357 R^{\rm AS}_{90}$. For the numerical results shown in this work, we will use axion star density profiles directly obtained from solving the equations of motions following Ref.~\cite{Visinelli:2017ooc}.\footnote{Note that we use the ``non-relativistic (single harmonic) limit'' of Ref.~\cite{Visinelli:2017ooc} and neglect axion self-interactions as appropriate for the ``dilute'' axion stars we are interested in here.}
 
For the velocity distribution of axions in the axion star we assume a flat distribution, $\tilde{f}(\vec{v}) \propto H(v - v_{\rm esc})$, truncated at the escape velocity $v_{\rm esc} = \sqrt{2 G M_{\rm AS}/R_{\rm AS}}$. Note that this yields a velocity distribution compatible with the uncertainty principle giving rise to the pressure support of axion stars. 

\section{From Gravitational Disruption to Photon Detection} \label{sec:grav_ph}

In general, the radio signal generated from axion-photon mixing is obtained by integrating the coupled equations of motion for the axion and photon over the trajectories of the in-falling axions. This problem simplifies significantly by noticing that the `non-resonant' mixing far from the neutron star 
(where $m_a \gg \omega_{\rm pl}$) scales with the power in the magnetic field at length scales comparable to the (inverse) momentum transfer ($\sim 1/m_a$)~\cite{Marsh:2021ajy}. For the axion masses we are interested in [$1/m_a \sim 1\,{\rm cm} \times \left( 26\,\mu{\rm eV}/m_a \right)$], this non-resonant contribution is negligible in neutron star magnetospheres, instead, the mixing is strongly dominated by local resonances near the neutron star which occur when the axion 4-momentum matches that of photons, $k_a^\mu \simeq k_\gamma^\mu$. Note that for non-relativistic particles in a strongly magnetized plasma, $k_a^\mu \simeq k_\gamma^\mu$ roughly corresponds to $\omega_p \simeq m_a$, where the plasma frequency $\omega_p = \sqrt{4\pi\alpha n_e / m_e}$, with $n_e$ being the electron/positron charge density, and $m_e$ being the mass. Focusing exclusively on the resonant contribution to the photon flux, one can express the photon production rate as 
\begin{equation}\label{eq:rate1}
    R_{a\rightarrow\gamma} = \int dS \, d^3 v \, \frac{f(\vec{r}, \vec{v})}{m_a} \,  |\vec{v} \cdot \hat{n}| \, P_{a\rightarrow \gamma} (\vec{r}, \vec{v}) \, ,
\end{equation}
where $S$ is the hyper-surface defined by the manifold over which $k_a^\mu = k_\gamma^\mu$, $\hat{n}$ is the normal to $S$, $f(\vec{r}, \vec{v})$ is the axion phase space (normalized to the energy density)\footnote{Note that we have defined $f(\vec{r}, \vec{v})$ using a notation that differs slightly from that of~\cite{Witte:2021arp}.}, and $P_{a\rightarrow \gamma}$ is the axion-to-photon conversion probability. It is worth noting the resonant condition $k_a = k_\gamma$ depends on the axion energy, the plasma frequency, and the relative angular orientation of the axion momentum with respect to the magnetic field; as a result of the angular dependence, the conversion surface is actually a fattened version of the 2-dimensional surface defined by $\omega_p = m_a$, with the width of the fattened volume at the level of $\lesssim 1 \, {\rm km}$. For simplicity, in what follows we neglect this fattening of the resonance hyper-surface, treating it instead as a 2-dimensional surface defined by $\omega_p = m_a$; we have verified that this approximation introduces a negligible error in the calculation.

The remainder of this section is devoted to describing in detail how we solve \Eq{eq:rate1}, and how geometric ray tracing methods can be used to relate \Eq{eq:rate1} to observable quantities such as the flux density. Our procedure relies on computing \Eq{eq:rate1} via a Monte Carlo (MC) integration, and thus the description below focuses primarily on how to draw, and subsequently weight, each of the MC samples.

\subsection{The Conversion Surface}

Let us start by focusing on the surface integral. We will assume throughout this work that the charge density of the magnetosphere is given by the charge-separated Goldreich-Julian (GJ) distribution $n_e \simeq 2 \vec{\Omega} \cdot \vec{B} / e$~\cite{goldreich1969pulsar}\footnote{Note that this expression neglects a relativistic factor which become important near the light cylinder $R_{\rm LC} = \Omega^{-1}$. The focus here, however,  is at distances $r \ll R_{\rm LC}$, and this term is negligible.}; here, $\vec{\Omega}$ is the angular velocity of the pulsar, and it is is assumed that $\vec{\Omega}$ is misaligned with respect to the magnetic field $\vec{B}$ (which we assume to be purely poloidal dipolar) by an angle $\theta_m$. Notice that definition of $n_e$ above is enough to uniquely describe the spatial structure of $\omega_p$, and thus we have also uniquely set the structure of the conversion surface itself.

In order to perform the surface integral in \Eq{eq:rate1}, we uniformly sample the conversion surface using the procedure described in~\cite{Witte:2021arp}. This allows us to re-express \Eq{eq:rate1} as
\begin{equation}\label{eq:rate2}
    R_{a\rightarrow\gamma} = \frac{1}{N} \sum_i 2\pi \mathcal{R}^2 \,  \int d^3 v \, \frac{f(\vec{r}_i, \vec{v})}{m_a} \,  |\vec{v} \cdot \hat{n}_{\vec{r}_i}| \, P_{a\rightarrow \gamma} (\vec{r}_i, \vec{v}) \, ,
\end{equation}
where the summation runs over the MC samples obtained at positions $\vec{r}_i$, and the factor $\mathcal{R}$ is the maximum radial distance chosen in the surface area scheme (\ie, $\mathcal{R}$ is chosen to be any number greater than the maximal radial distance of the conversion surface, see \eg\cite{Witte:2021arp}). The surface normal $\hat{n}_{\vec{r}_i}$ is obtained by taking the gradient of the plasma frequency at $\vec{r}_i$. 

Before continuing, it is worth emphasizing the charge-separated GJ model is only expected to provide a rough estimate of the plasma frequency of active pulsars (although it is in excellent agreement with the electrosphere model of dead pulsars at small radii, see \eg~\cite{Safdi:2018oeu}), and as such caution should be taken in the quantitative interpretation of the  results presented.

\subsection{The local axion phase space distribution}

We now turn our attention to the  integration over the local velocity distribution in \Eq{eq:rate2}, which is complicated by the fact that axion clumps falling through the magnetosphere will have their phase space `spaghettified', \ie it will become heavily concentrated along a narrow set of in-falling trajectories; this effect is illustrated (along with the characteristic scales in the problem) in \Fig{fig:spag}. In order to simplify the problem, we use Liouville’s theorem to relate the local phase space density to the phase space density far away from the neutron star. For practical purposes, we choose to work under the assumption that the disruption of a minicluster or axion star can be treated as an instantaneous event occurring at the Roche radius (this is the distance at which the tidal force exerted by the neutron star exceeds the self gravity of the object itself $r_R = R_{\rm AC} \left( 2 M_{\rm NS} / M_{\rm AC }\right)^{1/3}$).  This approximation implies that miniclusters and axion stars are treated as a freely propagating bodies at distances $r > r_R$, and the axions comprising these objects are treated as independent free-falling objects at $r < r_R$; see \eg Ref.~\cite{Bai:2021nrs} for a justification of this approximation. Switching the integration variable to the velocity at the Roche radius $v_{r_R}$, \Eq{eq:rate2}
can be expressed as 
\begin{multline*}
    \label{eq:rate3}
    R_{a\rightarrow\gamma} =  \frac{1}{N} \sum_i 2\pi \mathcal{R}^2 \,  \int d^3 v_{r_R} \, |{\bf J}(\vec{v})| \frac{f_{r_R}(\vec{r}_R[\vec{r}_i], \vec{v}_{r_R})}{m_a} \\
     \times |\vec{v} \cdot \hat{n}_{\vec{r}_i}| \, P_{a\rightarrow \gamma} (\vec{r}_i, \vec{v}) \, , 
\end{multline*}
where $|{\bf J}(\vec{v})|$ is the Jacobian relating the local velocity to that at the Roche radius, and $\vec{r}_R[\vec{r}_i]$ is the spatial position at the Roche radius consistent with producing a trajectory from $\vec{v}_{r_R}$ that intersects the conversion surface at $\vec{r}_i$. 

We can perform the velocity integral by drawing MC samples from the normalized velocity distribution at the Roche radius $f_{r_R}(\vec{v}_{r_R}) \equiv \tilde{f}_{r_R}(\vec{v}_{r_R} - \vec{v}_{AC})$, where we have boosted the velocity distribution discussed in Sec.~\ref{sec:mini_stars} by the relative velocity between the axion clump and the neutron star.  Neglecting relativistic effects and assuming conservation of energy and angular momentum, one can relate the $v_{r_R}$  to the local velocity  at a position $\vec{r}$ via~\cite{Alenazi:2006wu}
\begin{equation}\label{eq:velInf}
    \vec{v}_{r_R}[\vec{v}] = \frac{v_{r_R}^2 \vec{v} + v_{r_R} (G M_{\rm NS}/r)\hat{r} - v_{r_R} \vec{v}(\vec{v} \cdot \hat{r})}{v_{r_R}^2 + (G M_{\rm NS}/r) - v_{r_R} (\vec{v} \cdot \hat{r})} \, .
\end{equation}
For a fixed value of $\vec{v}_{r_R}$ and $\vec{r}$, this equation can be inverted to solve for a maximum number of two solutions $\vec{v}_i$. Under this procedure, the production rate simplifies to
\begin{eqnarray}
    \label{eq:rate4}
    R_{a\rightarrow\gamma} =  \frac{1}{N} \sum_i \sum_{j=1,2} 2\pi \mathcal{R}^2  \, |{\bf J}(\vec{v}_{i,j})| \,  n_a(\vec{r}_R[\vec{r}_i]) \nonumber \\
     \times |\vec{v}_{i,j} \cdot \hat{n}_{\vec{r}_i}| \, P_{a\rightarrow \gamma} (\vec{r}_i, \vec{v}_{i,j}) \, ,
\end{eqnarray}
where the summation over $j=1,2$ accounts for all possible solutions to \Eq{eq:velInf}, and we have re-expressed the energy density in terms of the number density $n_a$.

The only remaining piece is to determine the axion number density at $\vec{r}_R[\vec{r}_i]$. This can be obtained by evolving each axion trajectory backward to the Roche radius (under the assumption of Newtonian gravity). At the Roche radius, the number density can be directly evaluated for any choice of density profile and impact parameter. We note that while tracing the axion trajectory, we treat the interior of the neutron star to be at constant density\footnote{Notice that if we made the point mass approximation, we could have bypassed back-tracing the axion by setting the initial velocity and distance (the later fixed to the Roche radius), and enforcing conservation of angular momentum. We have used this procedure to instead verify the accuracy of this back-tracing procedure. } and take the mass and radius of the neutron star to be $M_{\rm NS}= 1 \, M_\odot$ and $r_{\rm NS} = 10\,$km. It is worth highlighting that for maximally allowed axion-nucleon couplings, the mean free path of axions through nuclear matter is many orders of magnitude larger than the neutron star radius, and thus we need not be concerned with issues of absorption.

\subsection{Photon Production }\label{sec:photonprod}

The final ingredient required to evaluate the photon production rate is the axion-photon conversion probability, which we take from~\cite{Millar:2021gzs}\footnote{This equation is derived in Ref.~\cite{Millar:2021gzs} by writing down the modified version of Maxwell's equations to include the background axion field, and looking for a plane wave solution in the WKB limit (\ie the limit in which the axion momentum is larger than the first derivatives of the electric fields). The final expression for the `conversion probability' reflects the ratio of the energy density stored in outgoing propagating Langmuir-O modes relative to the incoming axion field. }:
\begin{equation}\label{eq:convP}
P_{a\rightarrow \gamma} \simeq \frac{\pi}{2} \Big(1 + \frac{\omega_p^4 \zeta^2 \cos^2\theta}{\omega^4}\Big) \Big(\frac{\omega g_{a\gamma\gamma} B \zeta}{k_a}\Big)^2 \frac{1}{\lvert \partial_s k_\gamma \rvert} \, .
\end{equation}
Here, we have introduced the factor $\zeta \equiv \sin\theta / (1 - \omega_p^2 \cos^2\theta / \omega^2)$
and defined
\begin{equation}
\partial_s \equiv \partial_{\hat{k}_{\parallel}} - (\omega_p^2 \zeta \cos\theta / \omega^2) \partial_{\hat{k}_{\perp}} \, ,
\end{equation}
where $\hat{k}_{\parallel}$ and  $\hat{k}_{\perp}$ point in the direction parallel and perpendicular to the axion momentum, and we adopt a sign convention such that $\hat{k}_{\perp} \cdot \hat{B} > 0$. In this work, we focus on the small coupling regime where $P_{a \to \gamma} \ll 1$, and employ the perturbative calculation shown in \Eq{eq:convP}. Depending on the properties of the neutron star, the conversion probability can become $\mathcal{O}(1)$ for axion-photon couplings as small as  $g_{a\gamma\gamma} \sim 0.5 \times 10^{-12} \, {\rm GeV}^{-1}$; in that case, the perturbative calculation is no longer correct, and the photon flux may be significantly modified -- see e.g.~\cite{Foster:2022fxn,Carenza:2023nck} for a discussion.  Importantly, this implies that the flux density estimates shown here cannot simply be re-scaled in order to estimate the sensitivity to the axion-photon coupling.

Before continuing, a comment on the conversion probability is in order. The derivation of \Eq{eq:convP} assumes that $(i)$ axions and photons travel on approximately straight trajectories over the ``conversion length'' $L_c \sim \sqrt{\pi / |\partial_s k_\gamma|}$, and $(ii)$ that variations in the background can be approximated by a linear expansion. The second of these can be treated by either truncating the conversion length at the scale over which these assumptions fail, or by keeping higher order terms in the expansion~\cite{Millar:2021gzs} -- for the examples of interest, this can be treated be ensuring the conversion length stays below $\sim$km scales. Photon refraction, on the other hand, can invalidate the assumption of straight trajectories on much smaller distance scales. At the moment, the extent to which this premature photon refraction modifies the conversion probability is unclear. Reference~\cite{Witte:2021arp} has proposed a procedure for truncating the conversion length when refraction becomes significant, leading to a maximally conservative estimate of the conversion probability -- this technique is sometimes called the `$L_c$-cut', or the de-phasing cut. In what follows, we will in most cases apply the de-phasing cut so as to avoid the potential over-estimation of the radio flux (and we will clarify explicitly when this is not applied).

\subsection{Photon Propagation}
Until now, we have focused on computing the rate of photon production at the resonant conversion surface. In order to understand the properties of the radio flux, one must connect this rate with the distribution and properties of photons far from the neutron star. This connection can be accomplished using geometric ray tracing methods, which follow the group velocities of the sourced electromagnetic modes as as they refract and reflect off the background plasma (in what follows, we will use the term `photons' to refer to the group velocities of excited electromagnetic modes); such algorithms have already proven invaluable in understanding the radio properties generated from axions near neutron stars~\cite{Leroy:2019ghm,Witte:2021arp,Battye:2021xvt,Foster:2022fxn,Noordhuis:2022ljw}. 

Given a dispersion relation $\omega(\vec{x}, \vec{k}, t)$, the ray tracing equations are
\begin{eqnarray}
 \frac{d\vec{x}}{dt} &=& \nabla_k \omega(\vec{x}, \vec{k}, t) \;, \\ 
  \frac{d\vec{k}}{dt} &=& -\nabla_x \omega(\vec{x}, \vec{k}, t) \;, \\
  \frac{d\omega}{dt} &=& \partial_t \omega(\vec{x}, \vec{k}, t) \;,
\end{eqnarray}
where the third equation controls the dispersive effect of the plasma, giving rise to line-broadening. For the highly magnetized environments of interest, we are interested solely in Langmuir-O (L-O) mode, as this is the only propagating mode that mixes with the axion (mode mixing is not expected to arise in these environments, see \eg~\cite{Witte:2021arp}). The dispersion relation of the L-O mode is given by 
\begin{equation}
    \omega^2 = \frac{1}{2}\left(k^2 + \omega_p^2 + \sqrt{k^4 + \omega_p^4 + 2 k^2 \omega_p^2 (1-2\cos^2\tilde{\theta})} \right) \, ,
\end{equation}
where $\tilde{\theta}$ is the angle between $\vec{k}$ and the magnetic field.

We propagate all photons to a sphere around the neutron star with radius equal to that of the light cylinder $R_{\rm LC}$\footnote{Note that we choose this distance for two reasons: $(i)$ it is sufficiently far from the neutron star that photon trajectories are to a good approximation parallel, and $(ii)$ the assumption of a GJ charge distribution and dipolar magnetic field breaks down. Neglecting uncertainties associated to the latter point, this choice has been shown to be robust~\cite{Witte:2021arp}.}, bin the photons in pixelated regions on the sky of angular area $d\Omega_i$, and compute the differential power via
\begin{equation}
    \frac{dP}{d\Omega}(\theta, \phi) = \frac{1}{d\Omega_i} \sum_{j \in {\rm pixel}} \, R_{a\rightarrow \gamma, j} \, \times \, E_j \, ,
\end{equation}
where $R_{a\rightarrow \gamma, j}$ is the  weight (defined by the contribution to \Eq{eq:rate4}) of each photon, and the sum is confined to photons whose final locations are included in the pixel of interest\footnote{Here, we neglect resonant cyclotron absorption, which can induce an $\mathcal{O}(1)$ suppression of the flux for neutron stars with large magnetic fields (see \cite{Witte:2021arp}). }. The energy $E_j$ is set by the sum of the asymptotic energy of the axion prior to in-fall with the energy shift due to plasma broadening, and thus naturally accounts for the redshifting of the photon as it escapes the gravitational potential.

The flux density observed by a telescope $T$ is then given by
\begin{equation}
    S_T(\theta, \phi) = \frac{dP}{d\Omega}(\theta, \phi) 
    \, \frac{1}{\delta f_T \, \times \, d_T^2} \, ,
\end{equation}
where $d_T$ is the distance to the neutron star and $\delta f_T$ is the bandwidth of the observation.  The central value $\bar{E}$ and characteristic width $f_\sigma$ (in units of axion mass) of the spectral line in each bin can also be computed via
\begin{eqnarray}
 \bar{E} &=& \frac{ \sum_{j \in {\rm pixel}} \, R_{a\rightarrow \gamma, j} \, E_j }{\sum_{j \in {\rm pixel}} \, R_{a\rightarrow \gamma, j} } \;, \\
 f_\sigma &=& \left[\frac{\sum_{j \in {\rm pixel}} \, (E_j - \bar{E})^2 \, R_{a\rightarrow \gamma, j}}{ \bar{E}^2 \sum_{j \in {\rm pixel}} \, R_{a\rightarrow \gamma, j}} \right]^{1/2} \, . \label{eq:fsig}
\end{eqnarray}
For the examples of interest, the characteristic shift in $\bar{E}$ is much less than the effect of line broadening, and thus we will assume that the spectral properties are entirely determined by the latter.

\begin{table}
\begin{tabular}{ cc } 
\hline \\ 
\begin{tabular}{ |c|c| } 
\hline
$m_a$ & $26\,\mu$eV  \\ \hline
$g_{a\gamma\gamma}$ & $10^{-14} \, {\rm GeV}^{-1}$  \\ \hline
$B_0$ & $1.6 \times 10^{14}\,$G \\ \hline
$P$ & $3.76\,$s \\ \hline
$\theta_m$ & $0.2\,$rad \\ \hline
$\theta_v$ & 0.0\,rad  \\ \hline
$\delta f_T$ & $10^{-5} \times m_a$  \\ \hline
$d_T$ & 1\,kpc \\
\hline
\end{tabular}  %
\hspace{.2cm}
\begin{tabular}{ |c|c| } 
\hline
$M_{\rm AMC}$ & $10^{-12}\,M_\odot$  \\ \hline
$R_{\rm AMC}$ & $1.86 \times 10^9\,{\rm km}$  \\ \hline
$M_{\rm AS}$ & $10^{-13}\,M_\odot$  \\ \hline
$R_{\rm AS}$ & $3905\,{\rm km}$  \\ \hline
$\rho_{\rm AMC}$ & NFW  \\ \hline
$b_{\rm AMC}$ & $10^8\,\hat{x}\,$km  \\
\hline
$b_{\rm AS}$ & $2 \times 10^3\,\hat{y}\,$km  \\ \hline 
$|v_{\rm AMC/AS}|$ & 100\,km/s  \\ 
\hline
\end{tabular} \\
\\
\hline 
\end{tabular}
\caption{ Fiducial model parameters used to generate radio signals. From top to bottom, left to right, these parameters are: the axion mass ($m_a$), the axion-photon coupling ($g_{a\gamma\gamma}$), the dipolar magnetic field strength at the neutron star surface ($B_0$), the neutron star rotational period ($P$), the neutron star misalignment angle ($\theta_m$),  the incoming angle of the axion clump with respect to the neutron star rotation axis ($\theta_v$), the bandwidth of the observation ($\delta f_T$), the distance of the neutron star from Earth ($d_T$), the mass of the minicluster ($M_{\rm AMC}$), the radius of the minicluster ($R_{\rm AMC}$), the mass of the axion star ($M_{\rm AS}$), the radius of the axion star ($R_{\rm AS}$), the density profile of the minicluster ($\rho_{\rm AMC}$), the impact parameter of the minicluster ($b_{\rm AMC}$), the impact parameter of the axion star ($b_{\rm AS}$), and the relative velocity between the axion clump and neutron star ($v_{\rm AMC/AS}$). }\label{tab:fid}
\end{table}

\section{Results} \label{sec:results}

\begin{figure}
    \centering
    \includegraphics[width=0.45\textwidth]{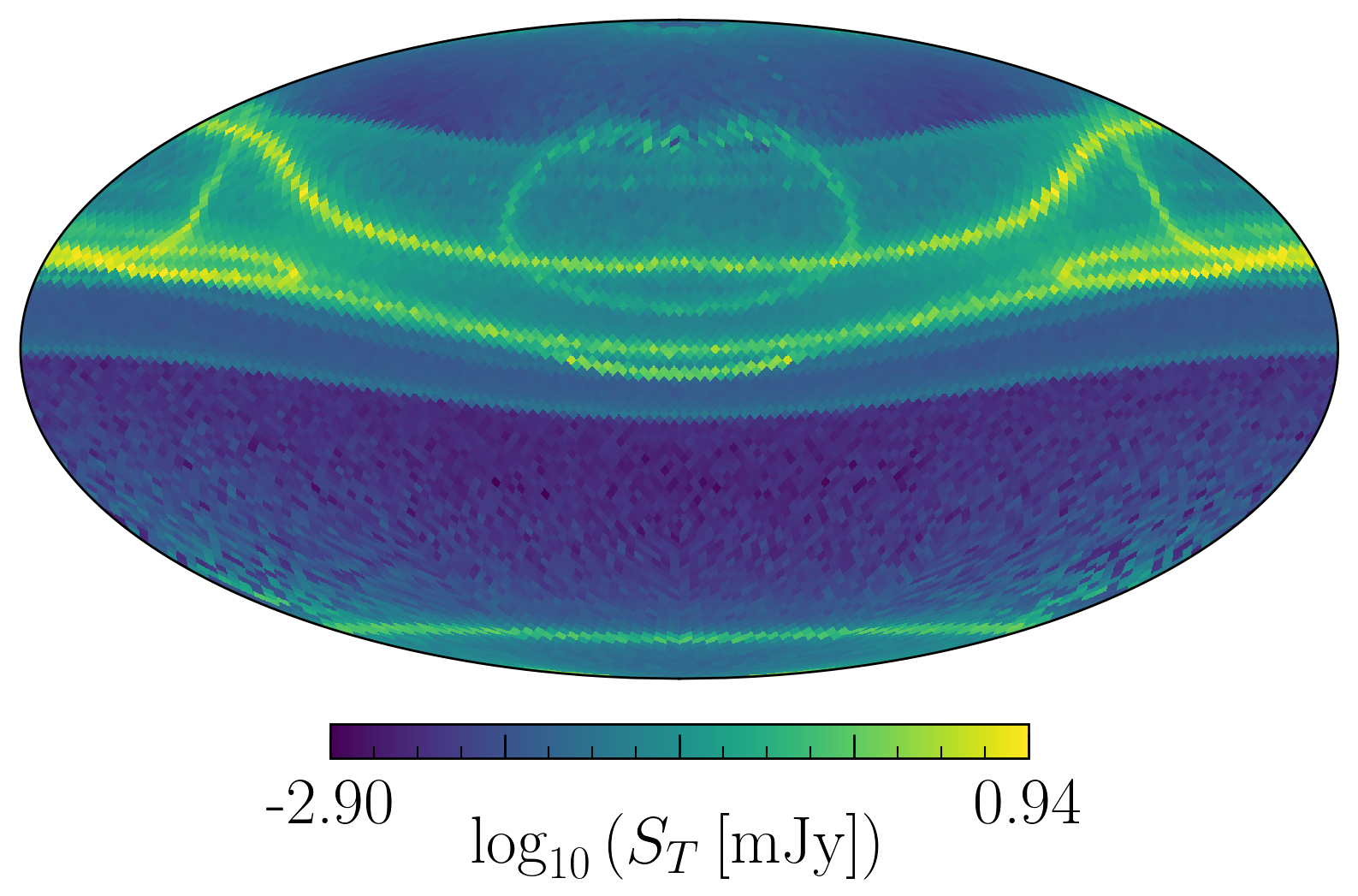}
    \includegraphics[width=0.45\textwidth]{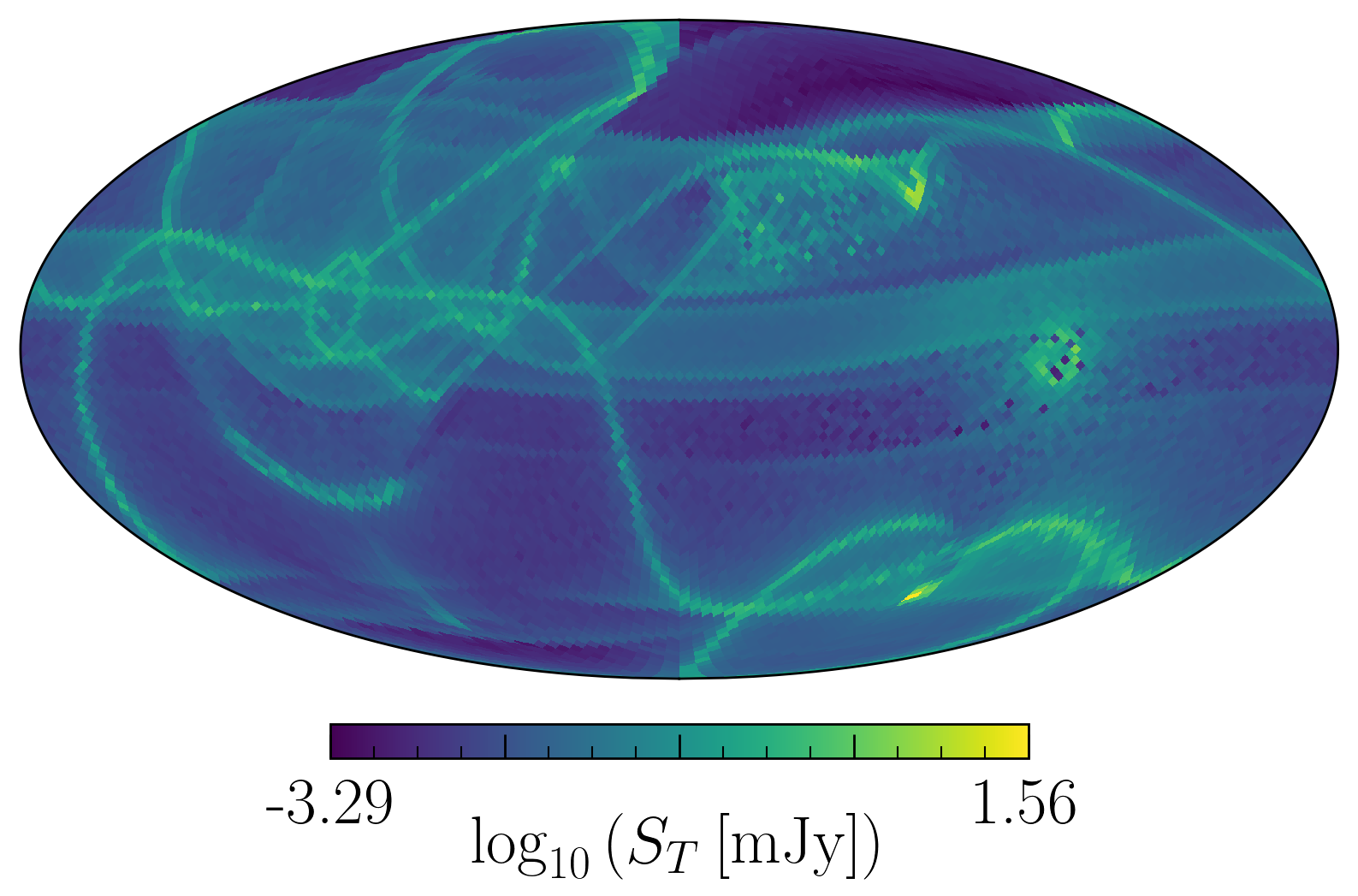}
    \includegraphics[width=0.45\textwidth]{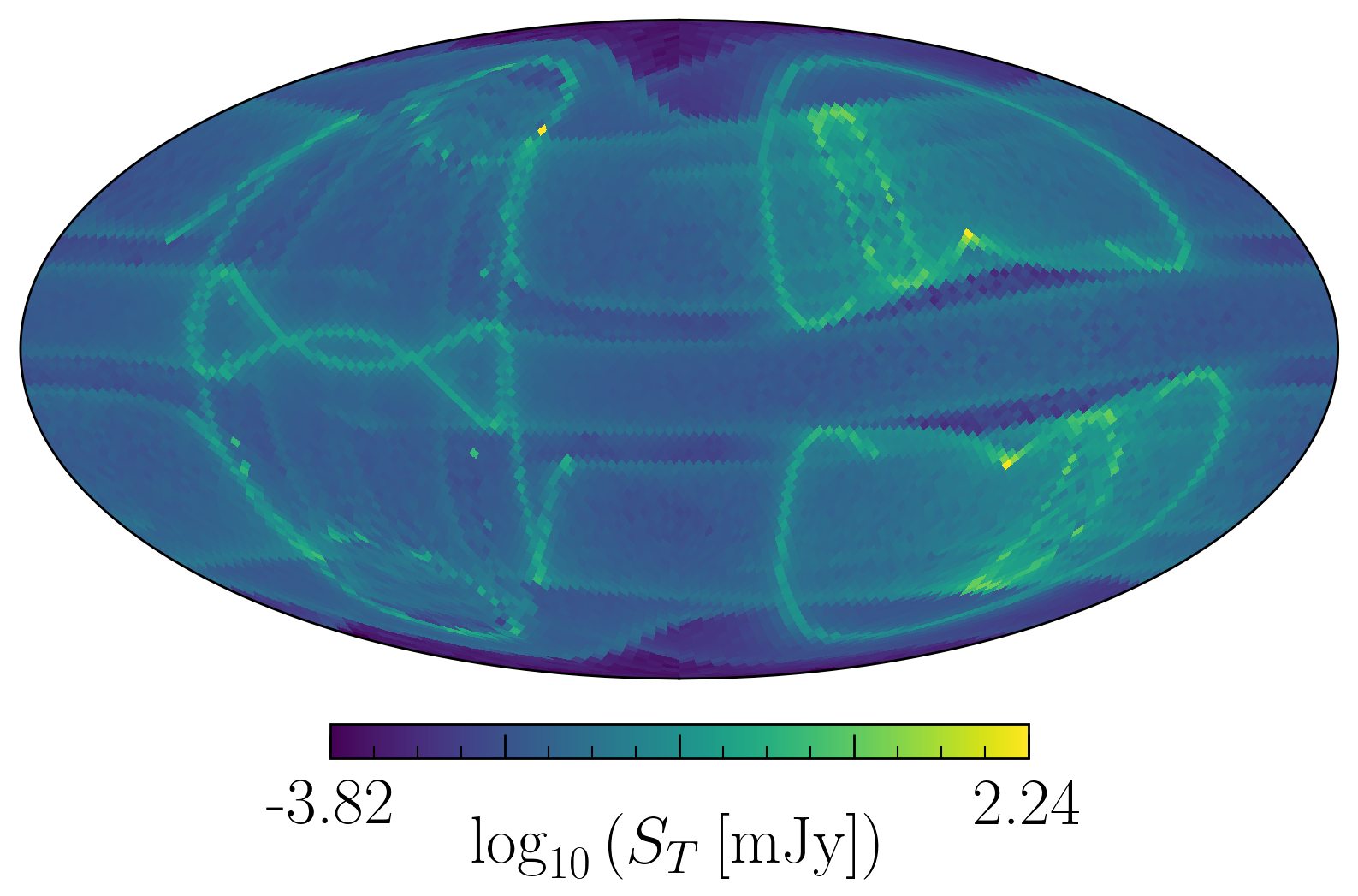}
    \caption{\label{fig:skymap_amc_nolc} Sky maps showing radio flux density $S_T$ arising from a minicluster-neutron star encounter. The different panels illustrate the effect of varying the encounter angle, defined by the relative angle between the minicluster velocity (in the neutron star rest frame) and the neutron star axis of rotation, with values of $\theta_v = 0$ (top), $\theta_v = \pi/4$ (center), and $\theta_v = \pi/2$ (bottom). All other parameters are set to the fiducial values outlined in \Tab{tab:fid}. Refraction induced de-phasing cut is not applied.}
\end{figure}

\begin{figure}
    \centering
    \includegraphics[width=0.49\textwidth, trim={0cm 1cm 0cm 2.2cm}, clip]{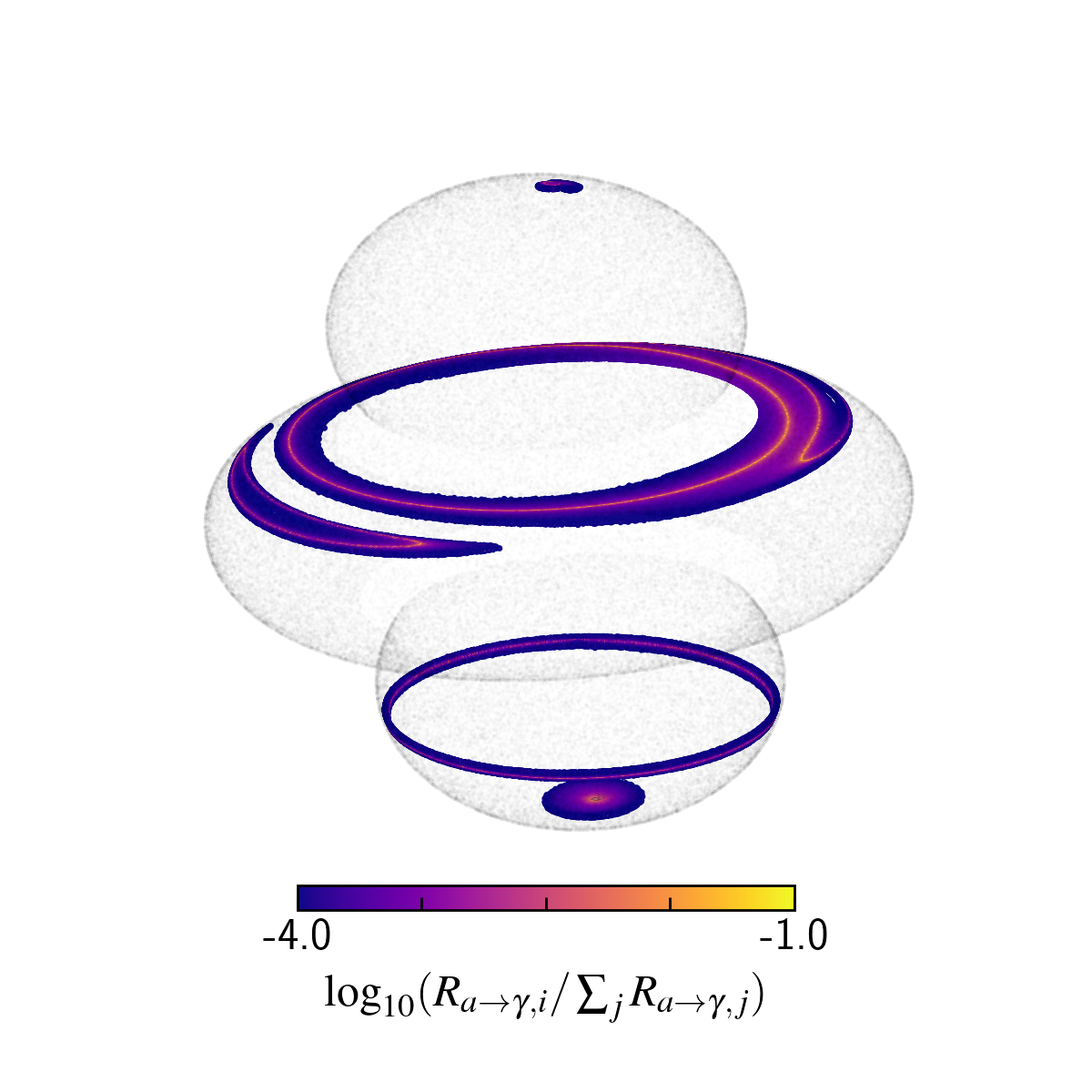} \\ \vspace{.3cm}
    \includegraphics[width=0.49\textwidth, trim={0cm 1cm 0cm 2.2cm}, clip]{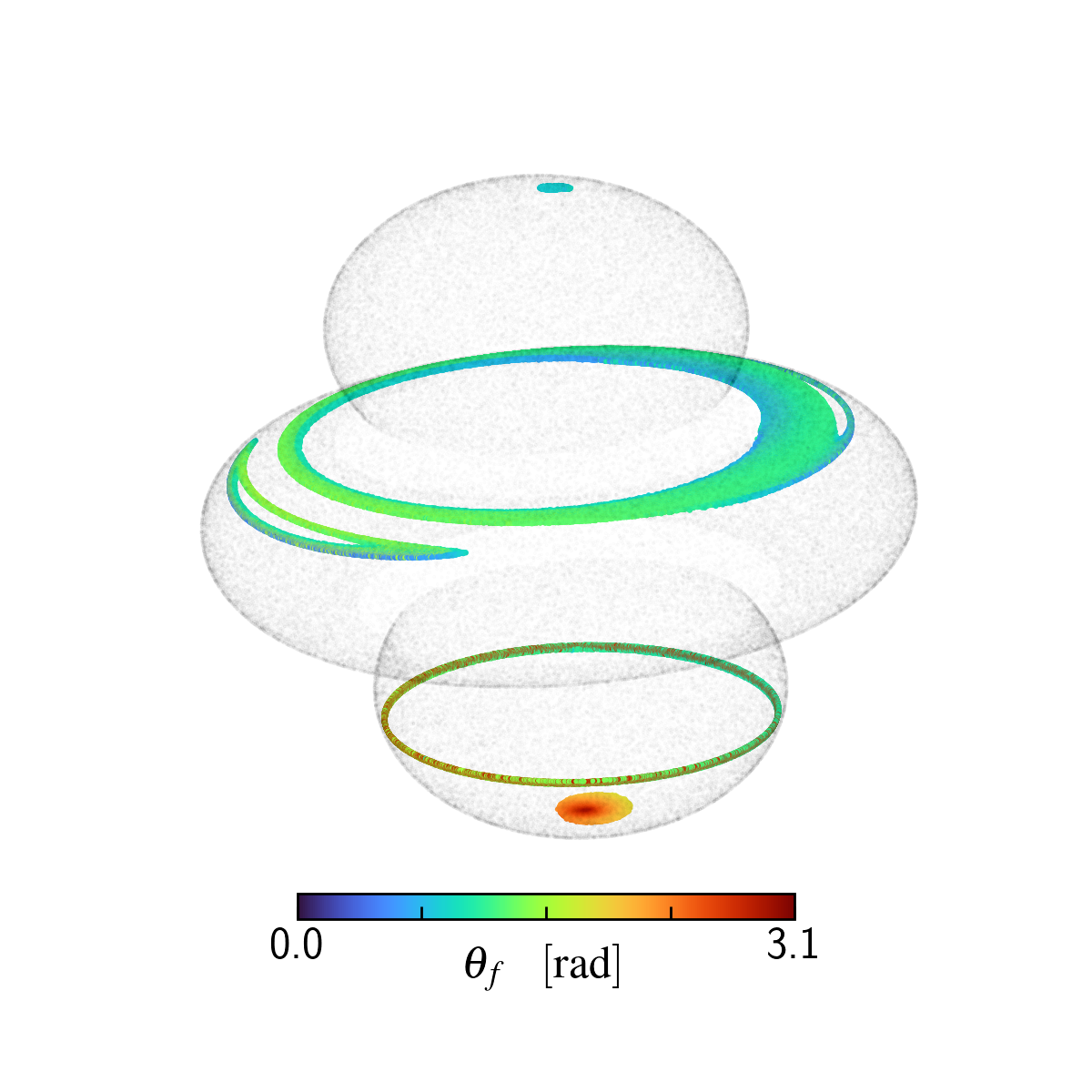}
    \caption{\label{fig:CS_proj_nolc} In order to understand the localized feature in the sky maps of \Fig{fig:skymap_amc_nolc}, we show where on the conversion surface the ``photons'' with the largest weights ($R_{a\rightarrow\gamma, i} / \sum_j R_{a\rightarrow\gamma, j} \geq 10^{-4}$) in our Monte Carlo sampling originate; recall that we are sampling photons uniformly over the surface area of the conversion surface. In the top panel, the color coding shows the relative contribution of these ``photons'' to the flux at the conversion surface in log base 10, ($\log_{10}(R_{a\rightarrow\gamma, i} / \sum_j R_{a\rightarrow\gamma, j})$, see Eq.~\eqref{eq:rate1}). In the bottom panel, the color coding indicates the ``photons'' final angular position on the sky $\theta_f$. Both panels are for the case of $\theta_v = 0$ (top panel of \Fig{fig:skymap_amc_nolc}).
    }
\end{figure}

\begin{figure}
    \centering
    \includegraphics[width=0.45\textwidth]{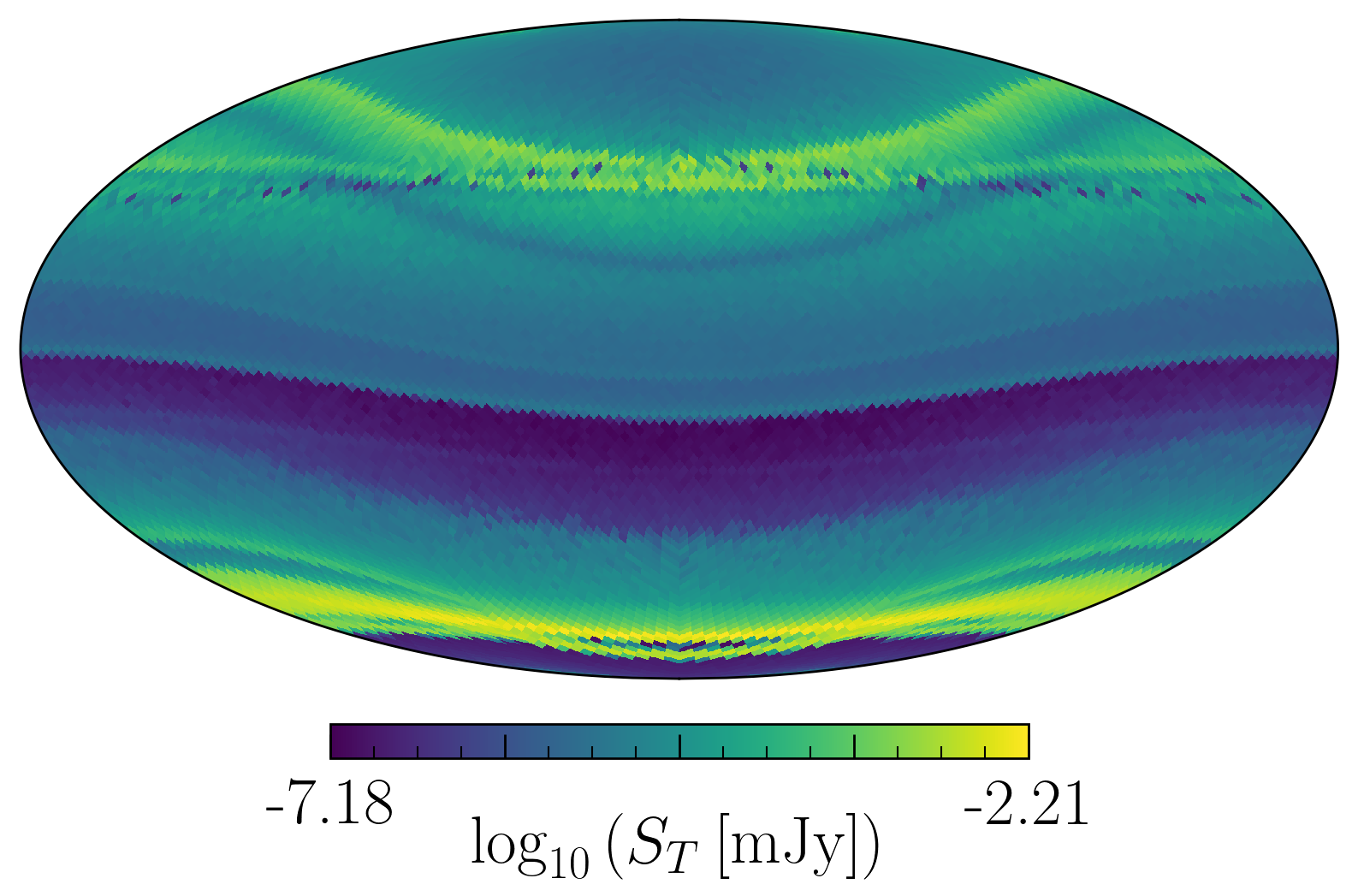}
    \includegraphics[width=0.45\textwidth]{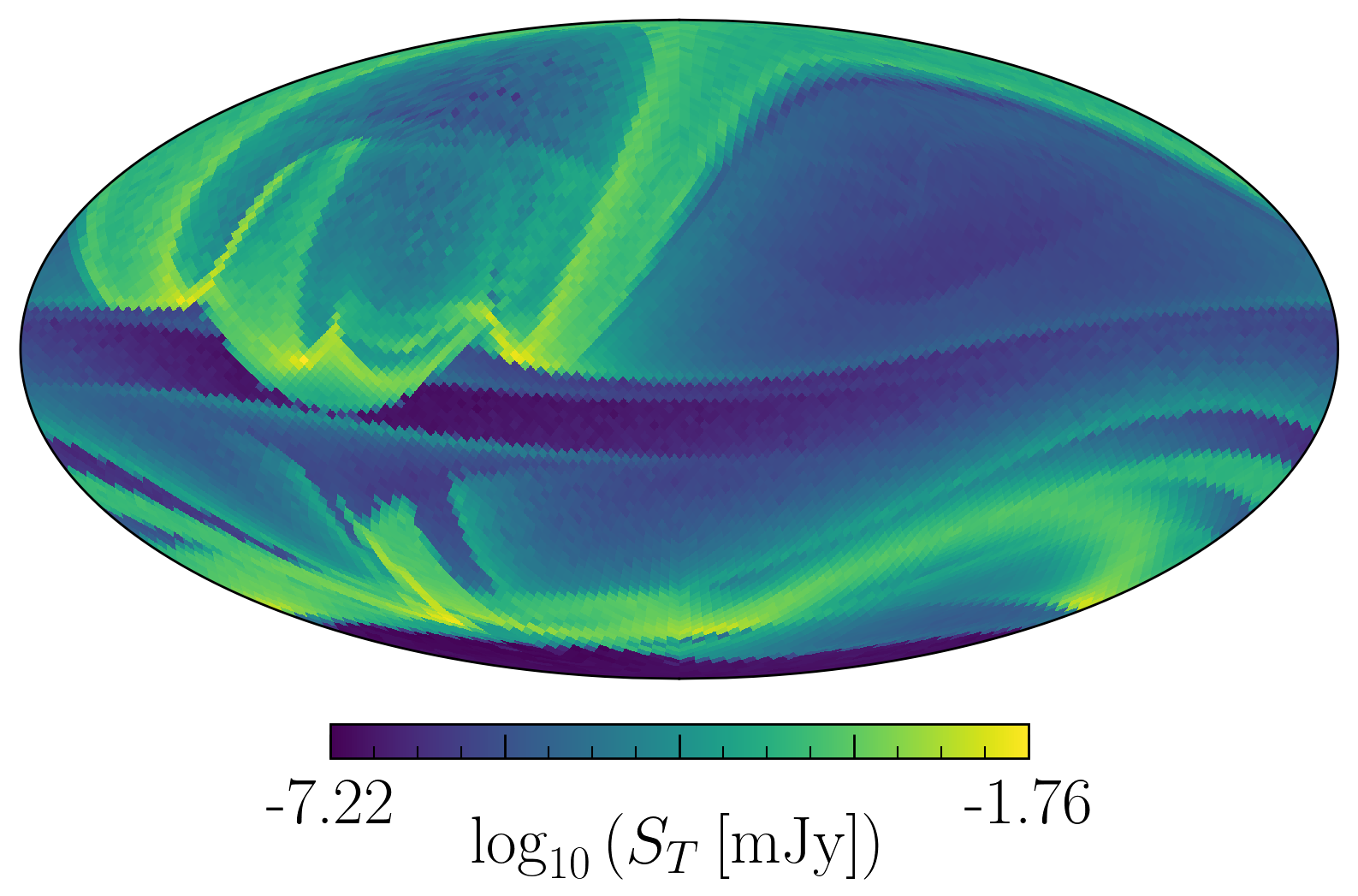}
    \includegraphics[width=0.45\textwidth]{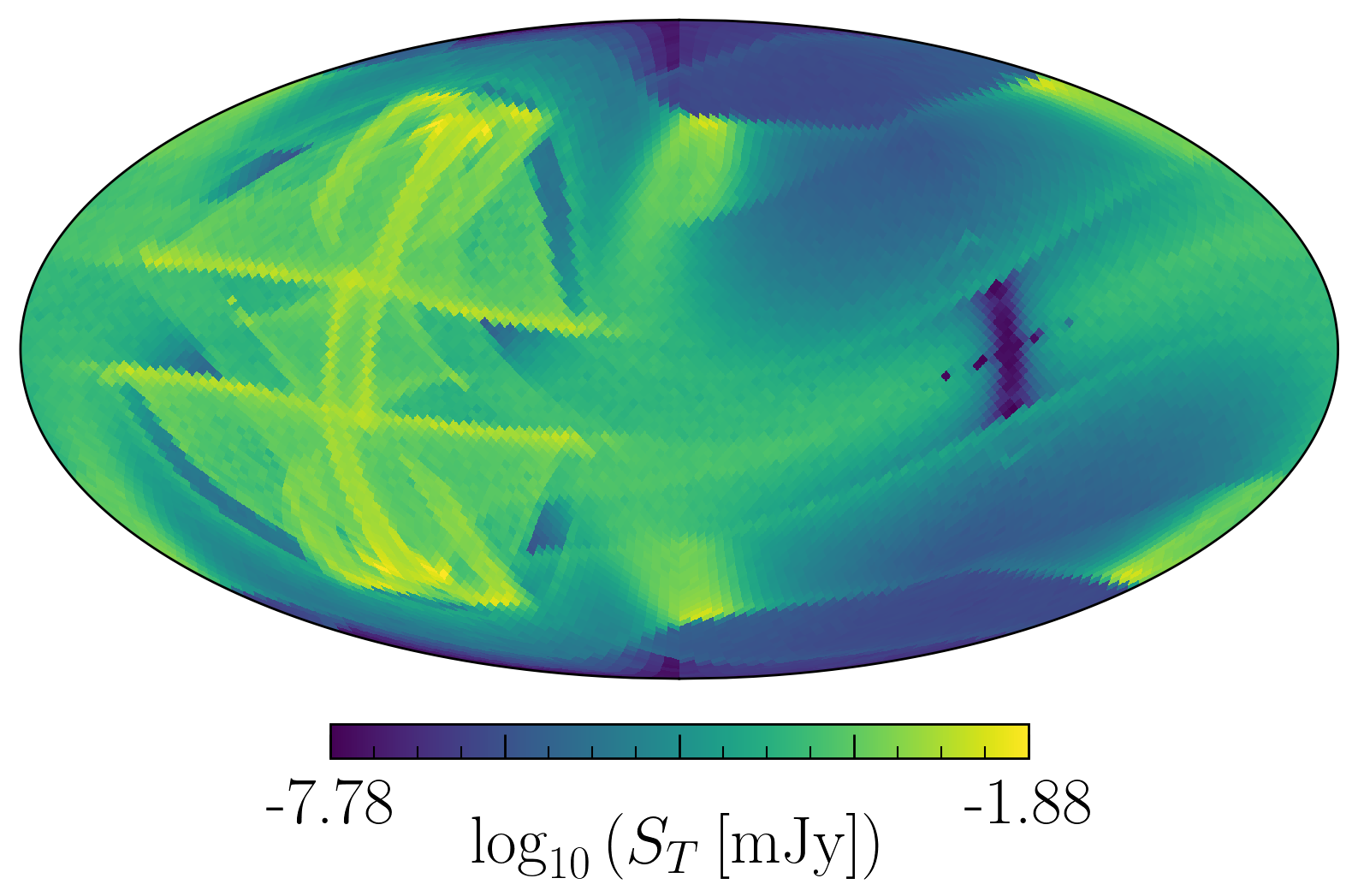}
    \caption{Same as \Fig{fig:skymap_amc_nolc} but applying the de-phasing cut on the conversion probability.}
    \label{fig:skymap_amc_lc}
\end{figure}

In this section we present the main results of this work, answering fundamental questions needed to search for axion clump-neutron star encounters, including: what is the expected magnitude and anisotropy of the radio transients, is there significant time-domain structure, what is the width of the spectral line, and how these properties change as a function of \eg impact parameter and relative velocity. In what follows, we adopt a fiducial set of parameters (listed in \Tab{tab:fid}), and vary them systematically in order to understand the impact of our assumptions.

\subsection{Anisotropy of Radio Flux}
Let us begin by looking at the anisotropy of the radio signal generated by a minicluster-neutron star encounter. In order to help disentangle the various effects that can contribute to the anisotropy of the flux, we begin by producing results without the de-phasing cut described in \Sec{sec:photonprod} (implying the flux densities are likely overestimated).

In \Fig{fig:skymap_amc_nolc} we plot the flux density observed across the sky (as viewed by an observer situated at the center of the neutron star) for various minicluster-neutron star encounters; the panels illustrate the effect of changing the orientation $\theta_v$ between the asymptotic minicluster velocity (in the neutron star rest frame) and the neutron star rotation axis. It is worth highlighting that all sky maps generated throughout the paper represent what an observer would view at a fixed snapshot in time; for simplicity, we choose this time to align with the peak of the flux density generated throughout the encounter (see \eg the time $t=0$ points in \Fig{fig:long_time}).  

\begin{figure}
    \centering
    \includegraphics[width=0.49\textwidth, trim={0cm 1cm 0cm 2.2cm}, clip]{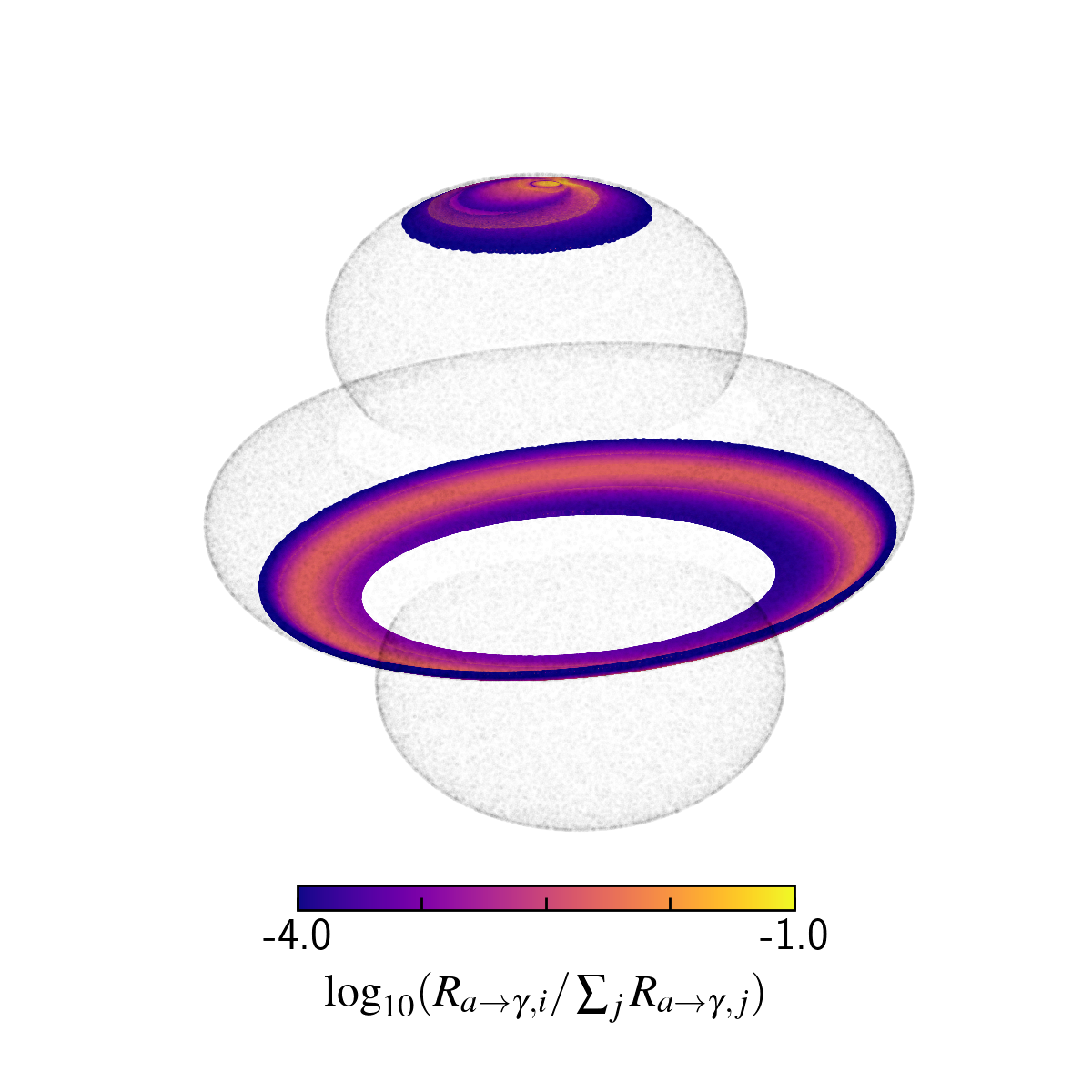} \\ \vspace{.3cm}
    \includegraphics[width=0.49\textwidth, trim={0cm 1cm 0cm 2.2cm}, clip]{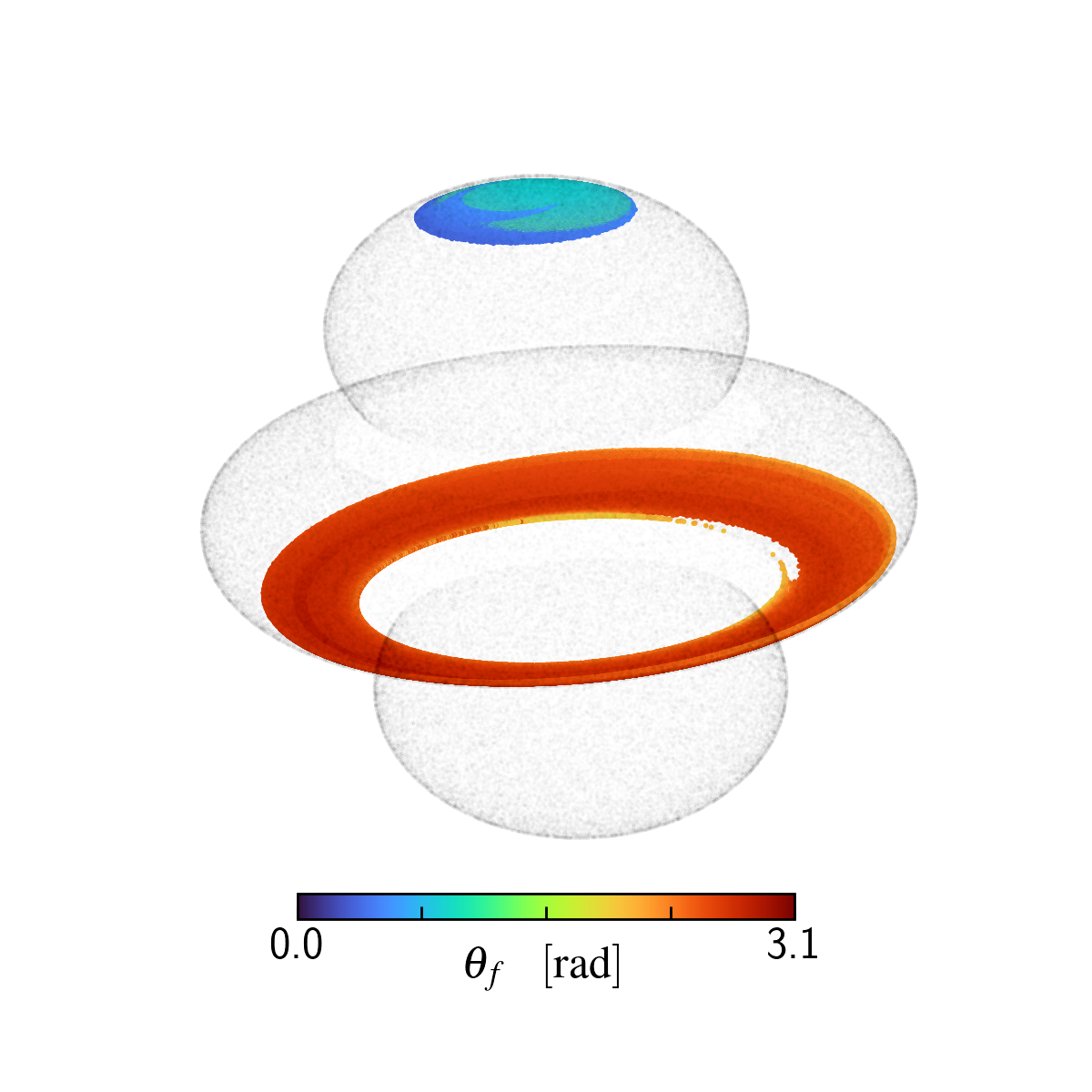}
    \caption{Same as \Fig{fig:CS_proj_nolc} but after applying the de-phasing cut on the conversion probability, \ie the $\theta_v = 0$ case shown in the top panel of \Fig{fig:skymap_amc_lc}.}
    \label{fig:CS_proj_lc}
\end{figure}

\begin{figure}
    \centering
    \includegraphics[width=0.45\textwidth]{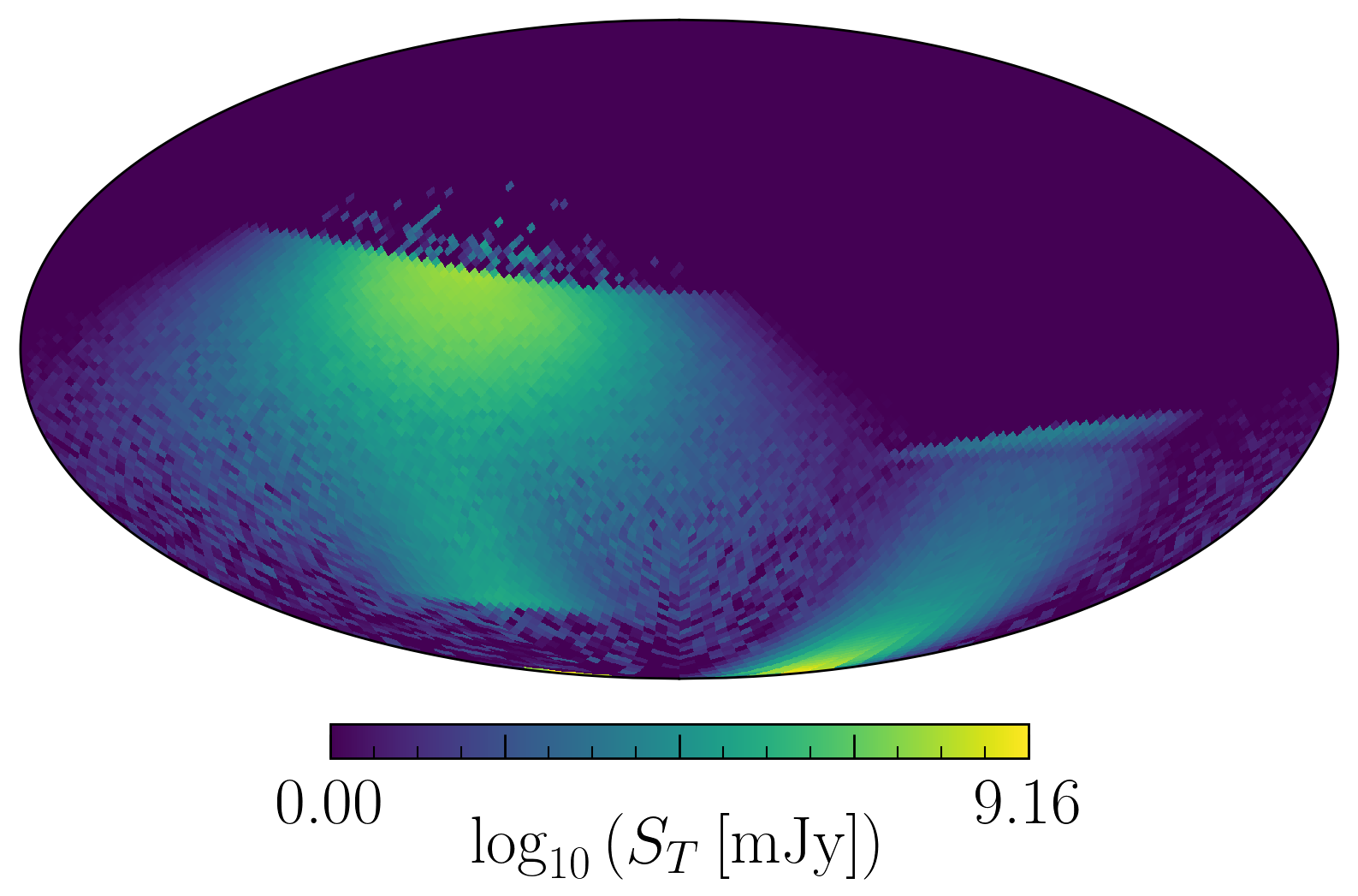}
    \includegraphics[width=0.45\textwidth]{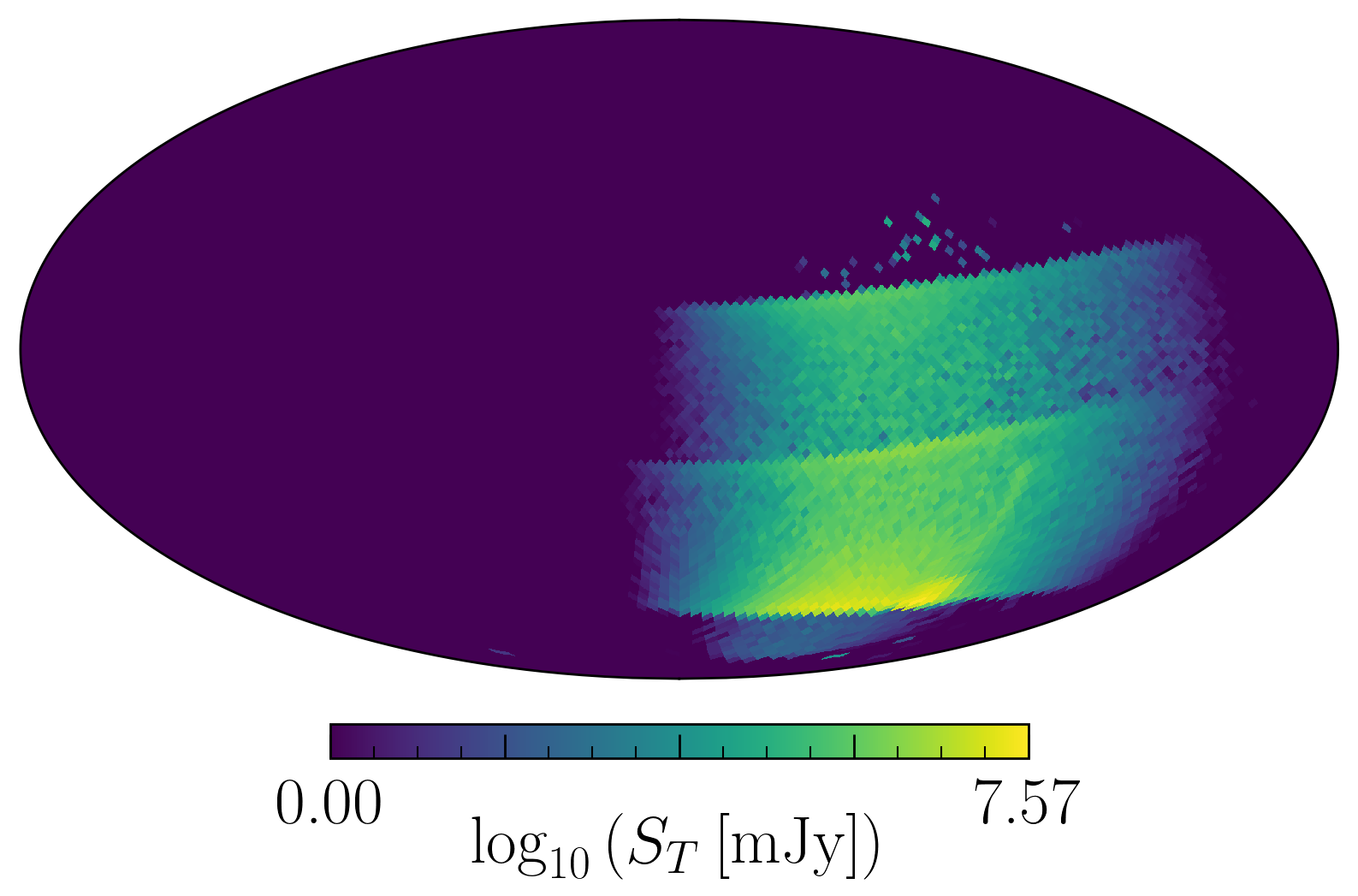}
    \includegraphics[width=0.45\textwidth]{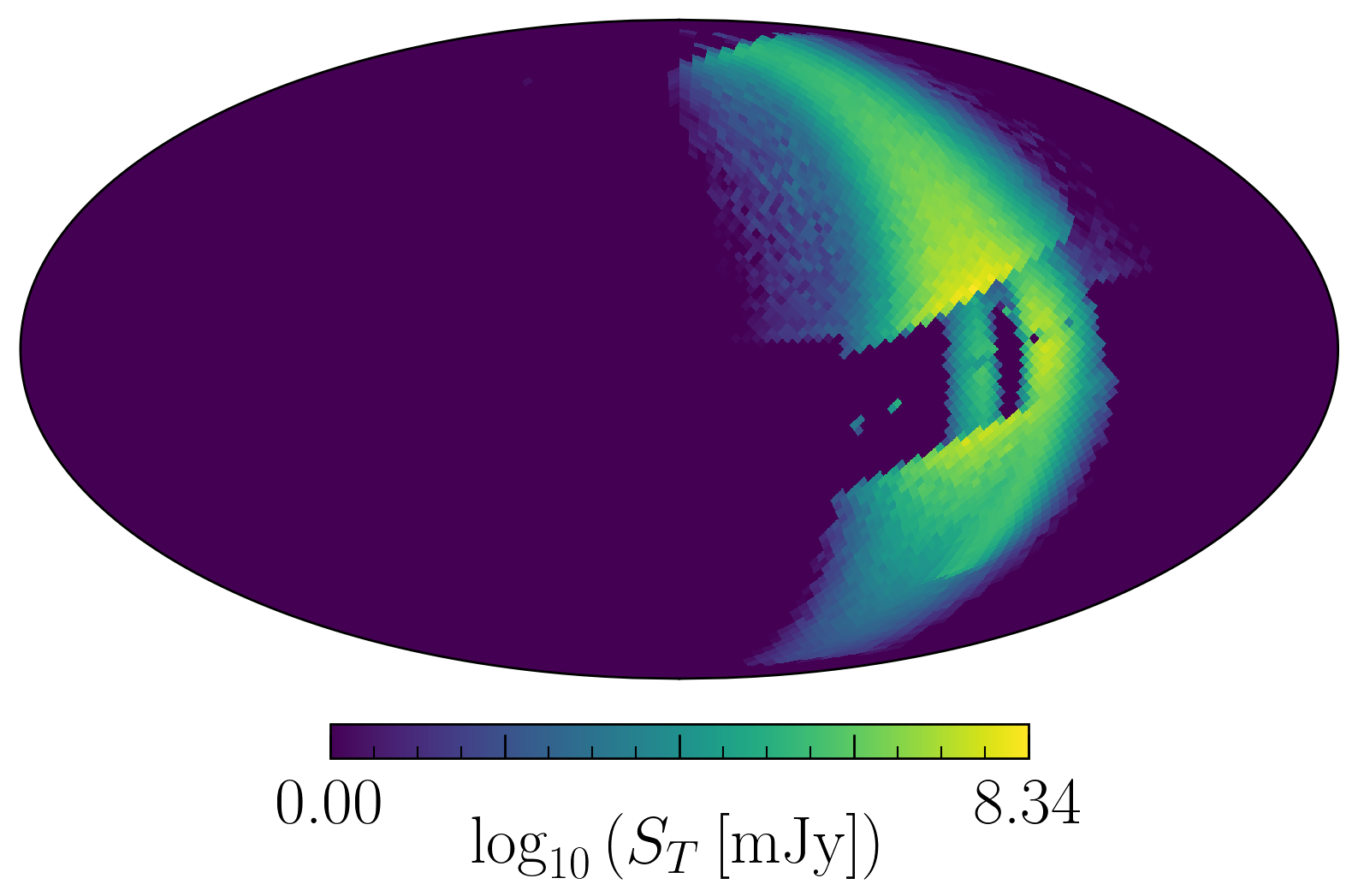}
    \caption{Same as \Fig{fig:skymap_amc_lc} but for an axion star-neutron star encounter.}
    \label{fig:skymap_as}
\end{figure}

\begin{figure}
    \centering
    \includegraphics[width=0.49\textwidth]{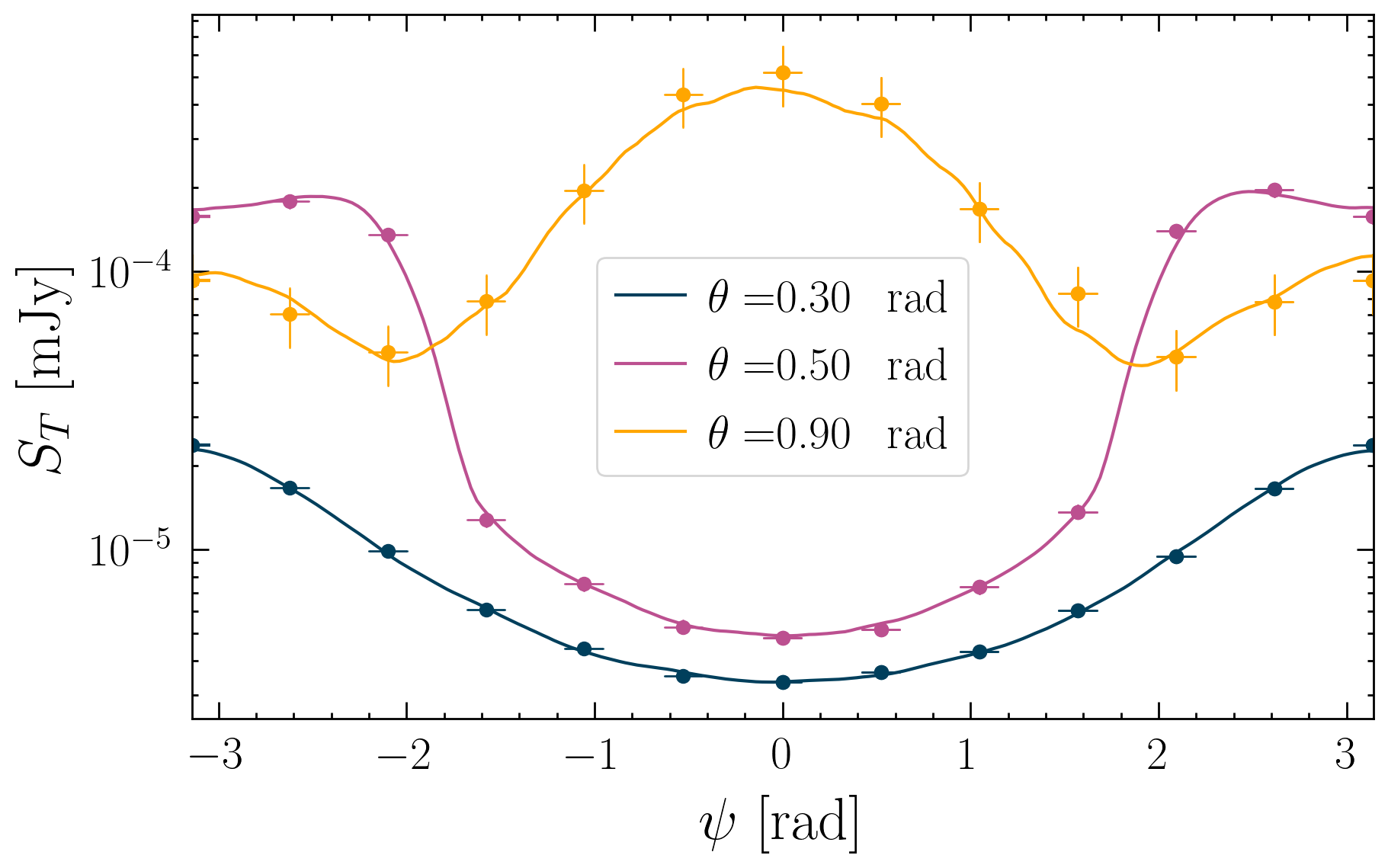}
    \includegraphics[width=0.49\textwidth]{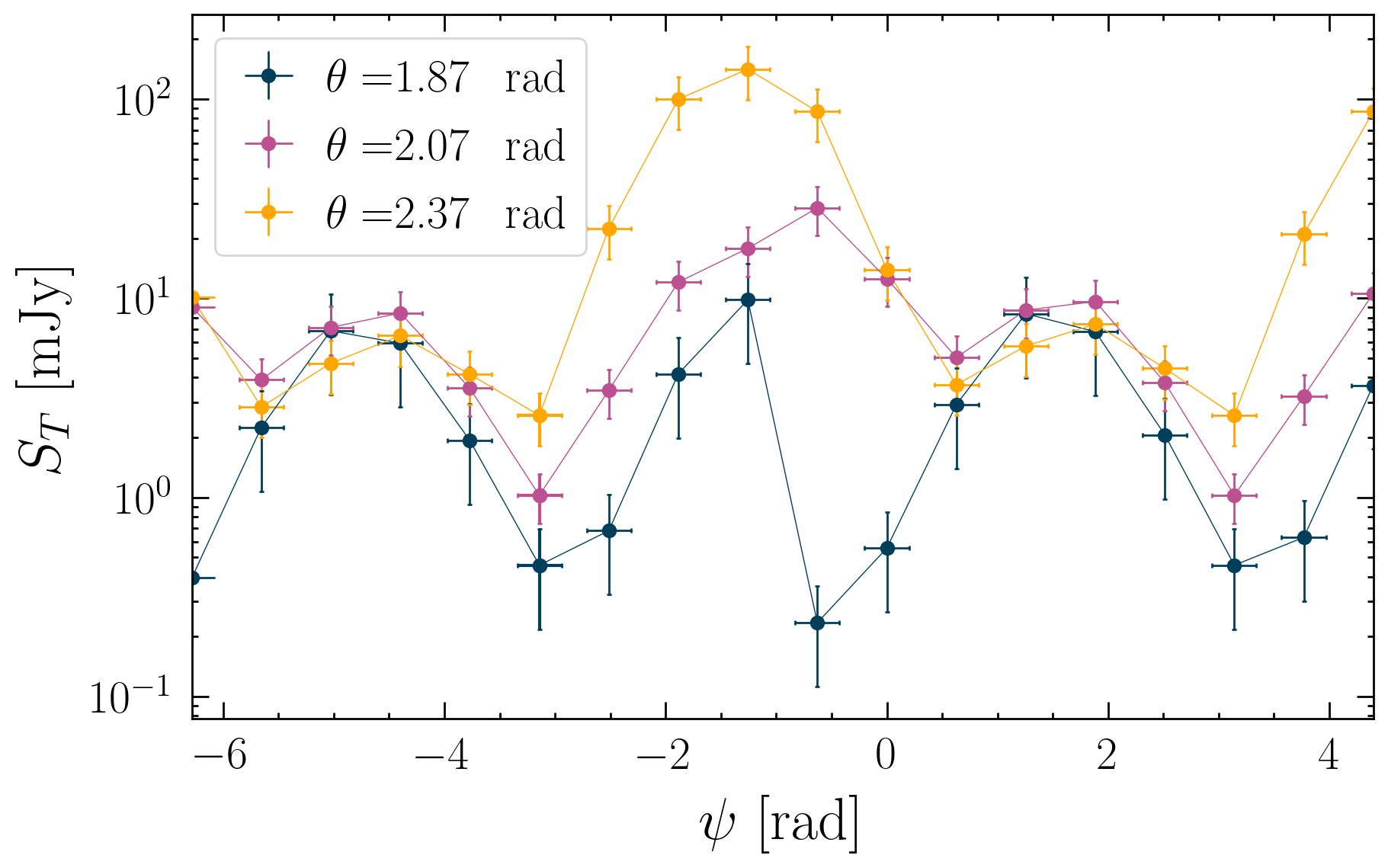}
    \caption{ Time evolution of the radio flux observed from a minicluster-neutron star (top) and axion star-neutron star (bottom) encounter. The differently-colored symbols/lines correspond to different viewing angles as denoted in the legend. The $x$-axis shows the rotational phase of the neutron star, $\psi \equiv 2 \pi t / P$, where $t$ is the time of an observer and $P$ is the rotational period of the neutron star. In both panels, the symbols show the average obtained from a set of Monte Carlo simulations performed for a fixed value of $\psi$, and the error bars show the variation of the results of the individual Monte Carlo runs (see text for details). In the top panel, the lines show the approximate time-evolution  that would be recovered by taking an azimuthal slice from the fixed-time sky map shown in \Fig{fig:skymap_amc_lc}. The lines in the bottom panel are merely an interpolation between neighboring points, and are intended to aid the reader in following the time evolution. }
    \label{fig:timeD}
\end{figure}

\begin{figure}
    \centering
    \includegraphics[width=0.45\textwidth]{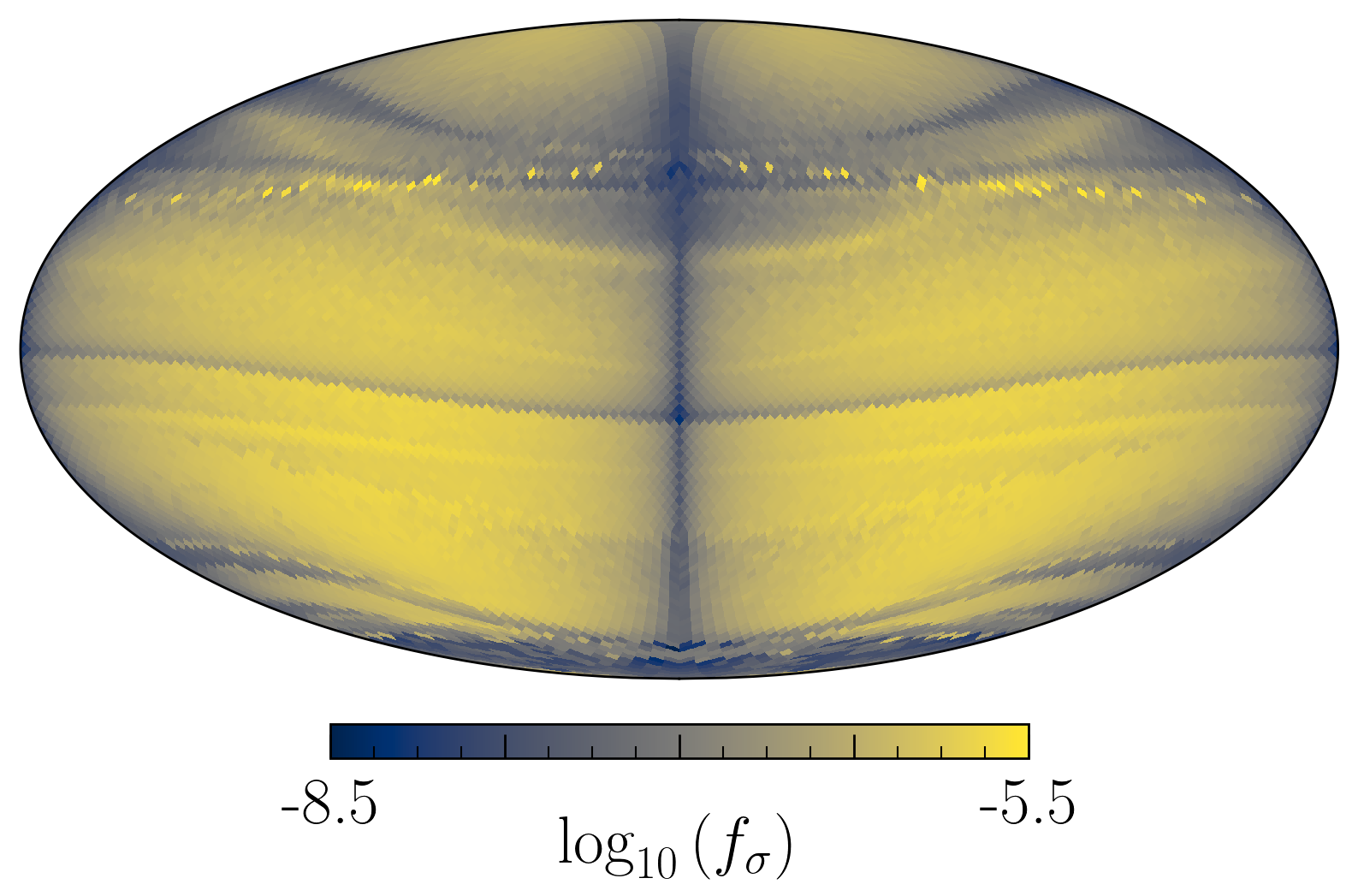}
    \includegraphics[width=0.45\textwidth]{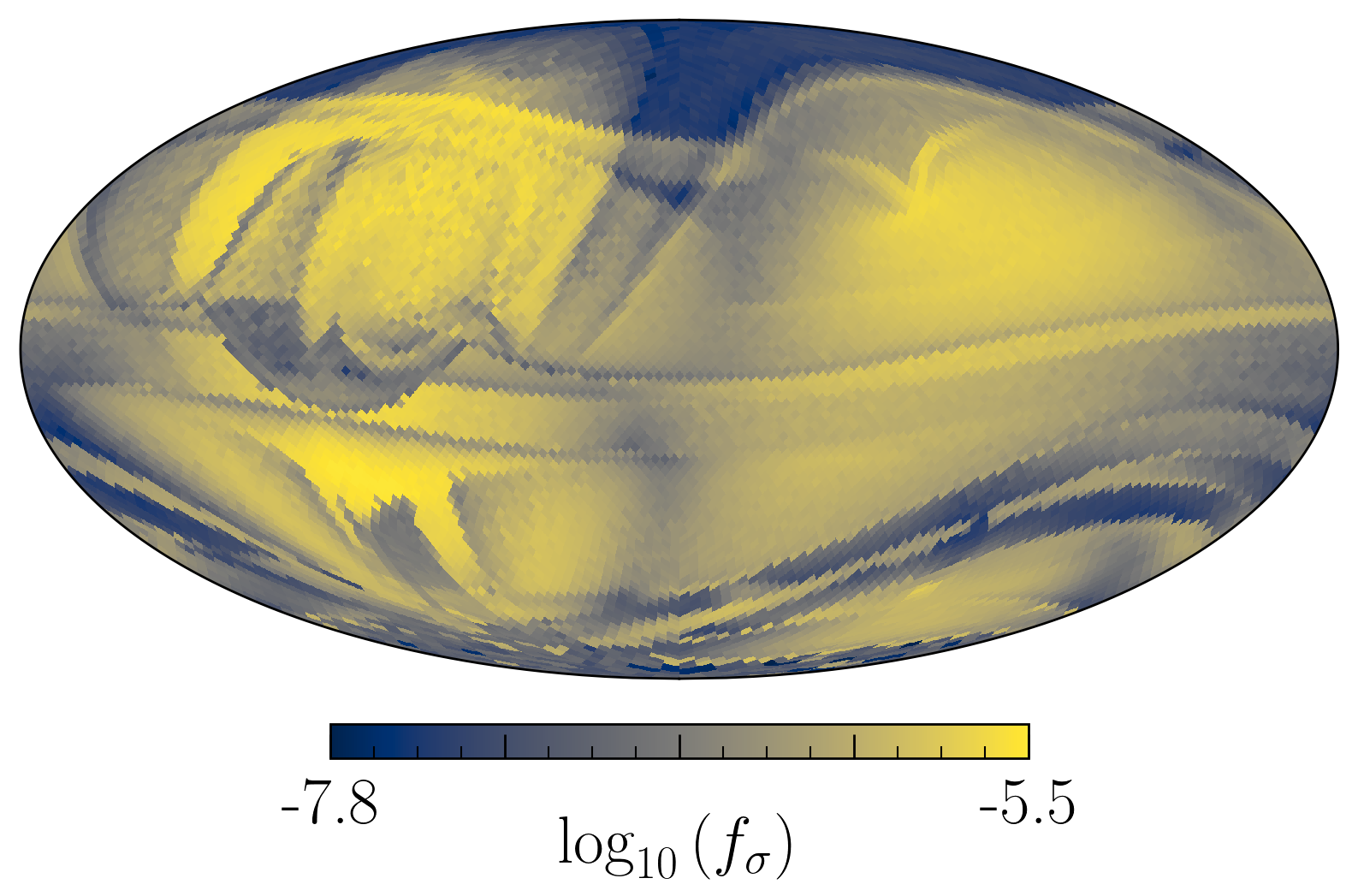}
    \includegraphics[width=0.45\textwidth]{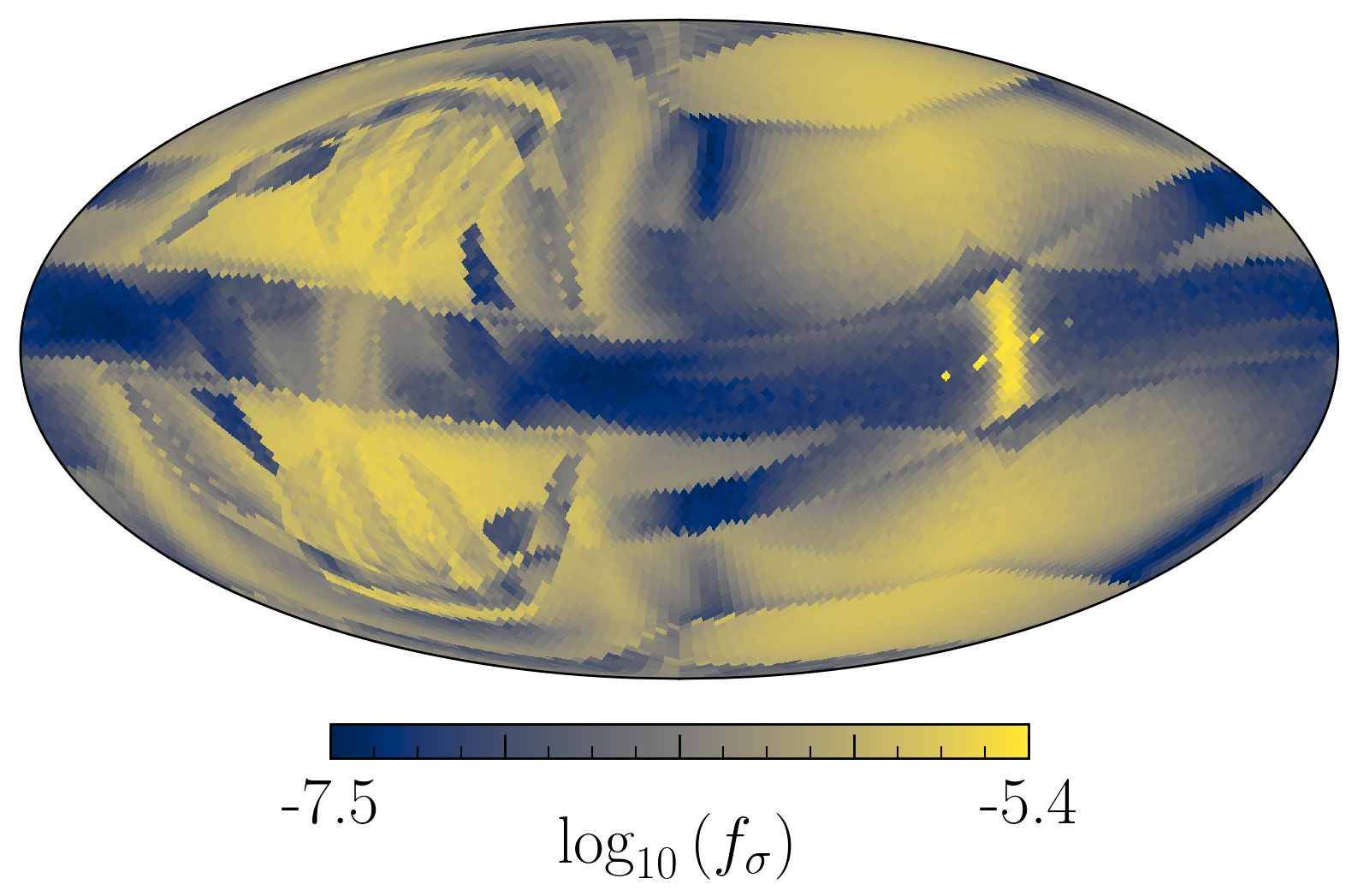}
    \caption{Same as \Fig{fig:skymap_amc_lc} (\ie for axion minicluster-neutron star encounters) but showing the characteristic width of the transient line (in units of the axion mass) $f_\sigma$.}
    \label{fig:skymap_width_amc_lc}
\end{figure}

\begin{figure}
    \centering
    \includegraphics[width=0.45\textwidth]{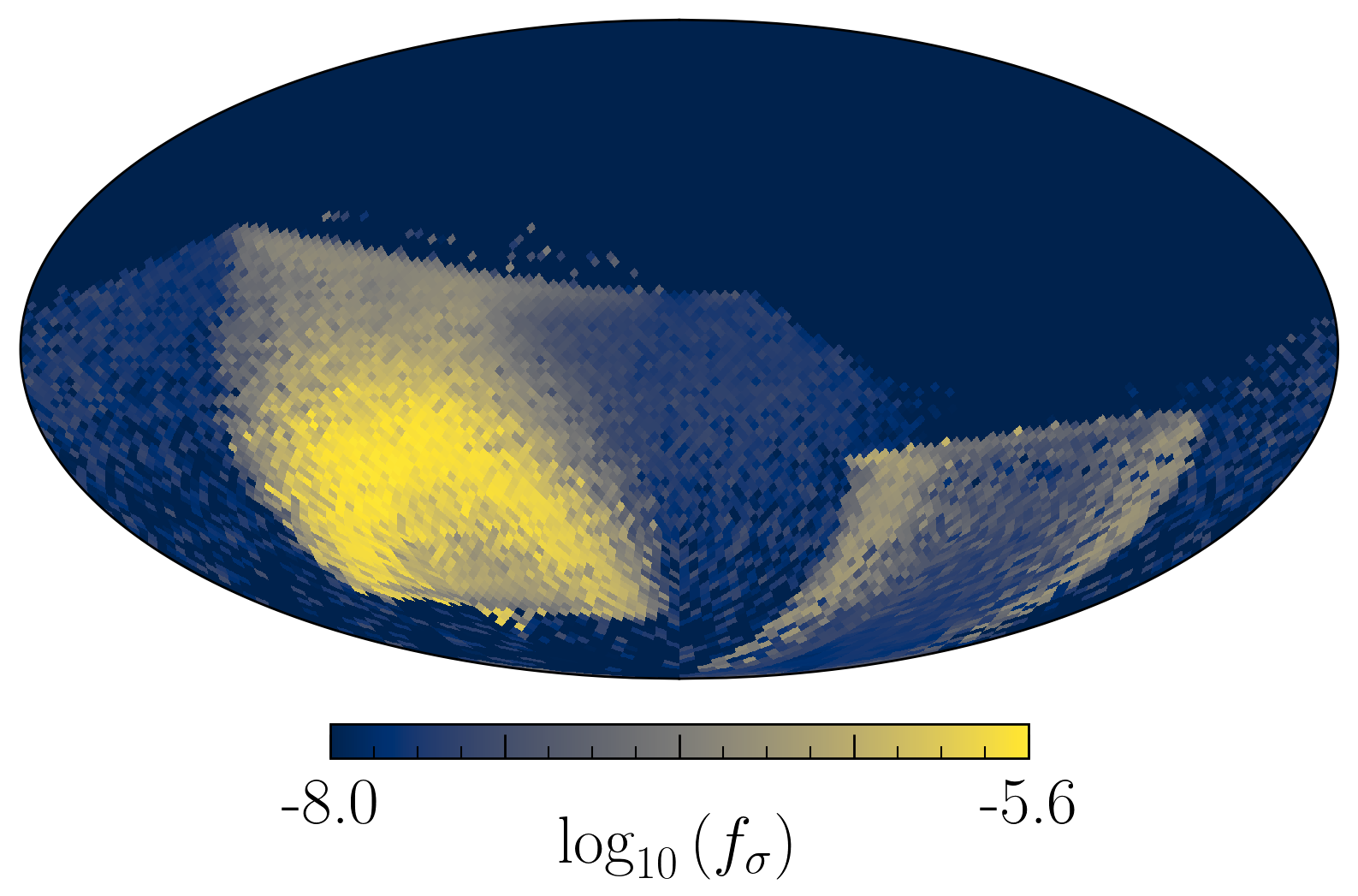}
    \includegraphics[width=0.45\textwidth]{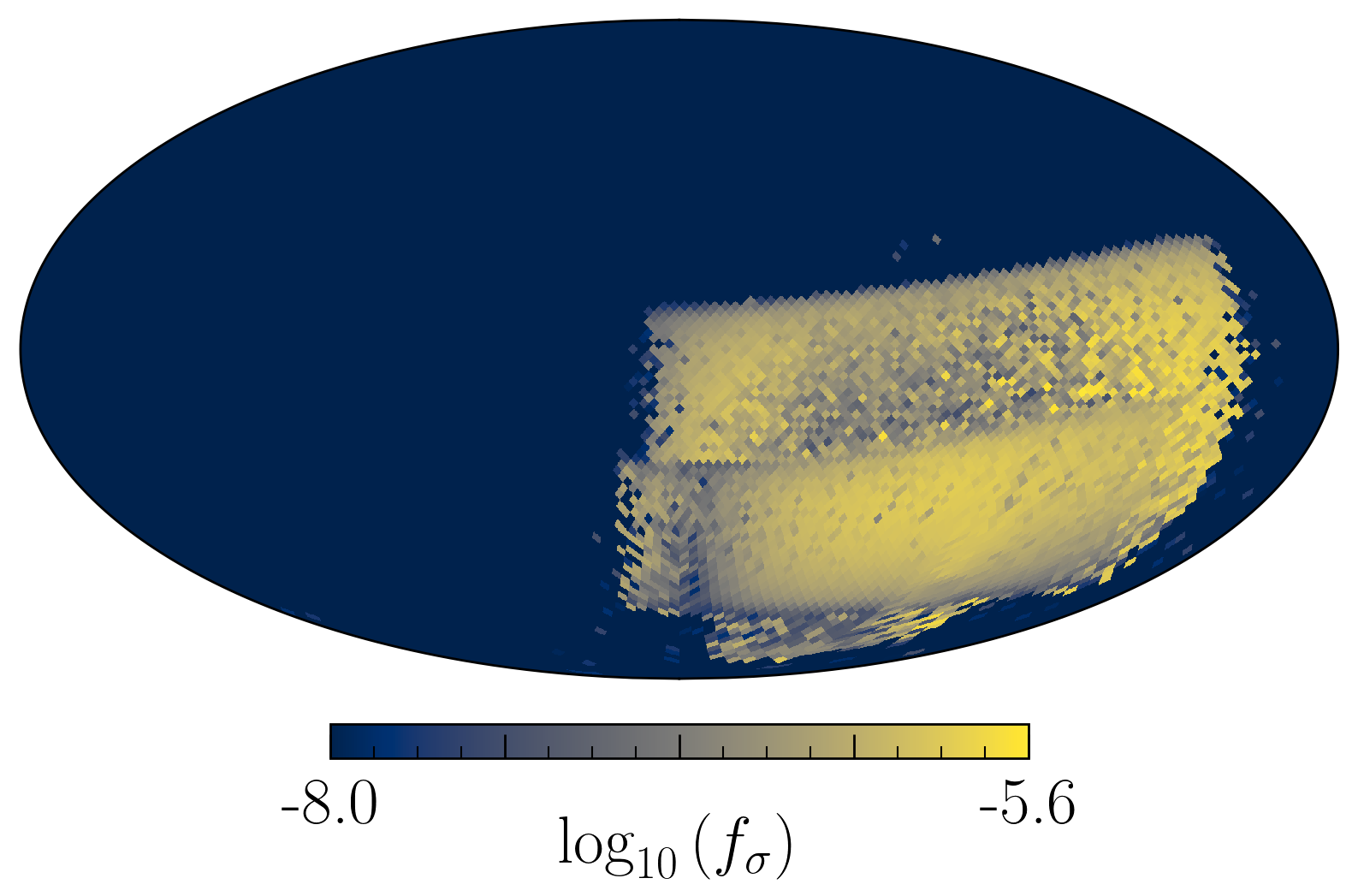}
    \includegraphics[width=0.45\textwidth]{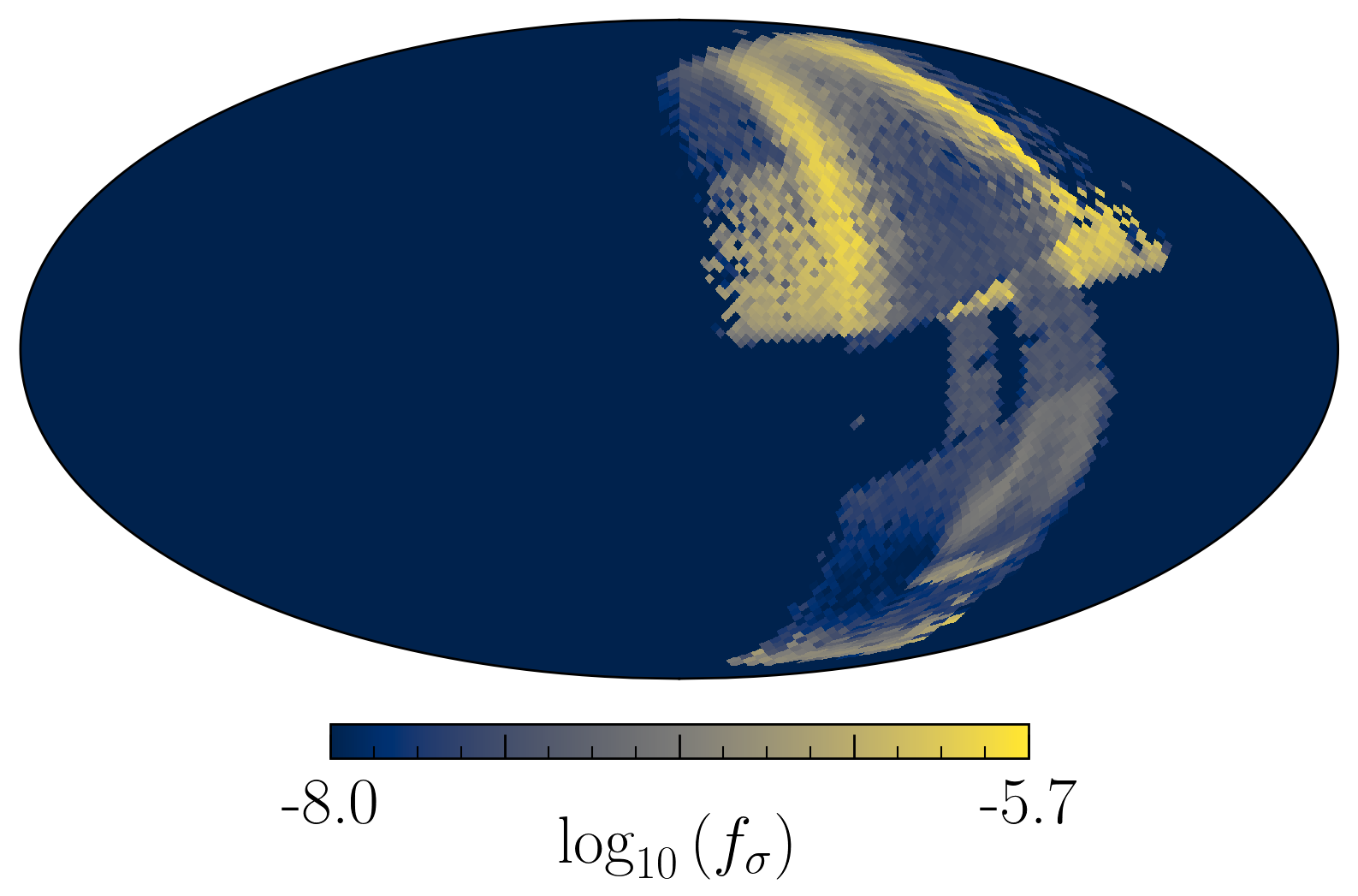}
    \caption{Same as \Fig{fig:skymap_as} (\ie for axion star-neutron star encounters) but showing the characteristic width of the transient line (in units of the axion mass) $f_\sigma$.}
    \label{fig:skymap_width_as}
\end{figure}

The maps presented in \Fig{fig:skymap_amc_nolc} show  complex small scale structures which are not easy to explain from the intrinsic geometry of the problem. In order better under their origin, we identify the photons in the $\theta_v=0$ map (top panel) for which $R_{a\rightarrow\gamma, i} / \sum_j R_{a\rightarrow\gamma, j} \geq 10^{-4}$, and project these photons back onto the conversion surface.  This procedure is illustrated in \Fig{fig:CS_proj_nolc}, where each photon has been colored by its relative contribution to the flux (top) or its final sky location $\theta_f$ (bottom; again, this angle $\theta_f$ is defined relative to the axis of rotation). The bright bands identified in \Fig{fig:skymap_amc_nolc} emanate from very small localized regions in the neutron star magnetosphere -- it is straightforward to identify these regions as arising from `glancing' axion trajectories, \ie trajectories for which $|\partial_s k_\gamma|$ is very small (note that this roughly corresponds to axion trajectories moving perpendicular to the plasma gradient at the conversion surface). In the case of the smooth dark matter halo, the relative phase space associated with such glancing trajectories is small, and thus such features do not arise; here, however, no such suppression exists; rather, the phase space is effectively `spaghettified', \ie it is heavily concentrated along a narrow set of in-falling trajectories (note that if the asymptotic velocity distribution were a pure delta function, each point on the conversion surface would be uniquely defined by at most two velocity vectors).

 In order to remain as conservative as possible in our estimates of the signal strength, we apply the de-phasing cut to all subsequent calculations. This de-phasing is driven by photon refraction, one expects a strong suppression in the conversion probability of glancing axion trajectories since the large tangential plasma gradients help drive pre-mature refraction. In order to visualize this effect, we repeat the above procedure for the same set of minicluster-neutron star encounters, but applying the de-phasing cut -- the results are shown in \Fig{fig:skymap_amc_lc} and \Fig{fig:CS_proj_lc}. One can see that the de-phasing cut induces both a net suppression of the flux density, a significant smoothing of all small-scale features, and a relative shift in the sky location of the peak flux. The projection plots also illustrate that the brightest photons now emanate from a much broader region across the conversion surface.

Owing to the compact nature of axion stars, axion star-neutron star encounters are expected to give rise to much stronger, but also more anisotropic, signals. We illustrate the magnitude and anisotropy that can arise from such encounters in \Fig{fig:skymap_as}, where as before we have shown the results for our fiducial model parameters (including the de-phasing cut), and varied the relative orientation of the encounter from $\theta_v = 0$ to $\theta_v = \pi/2$. Notice that at its brightest, the flux density  is increased by up to $\sim$11 orders of magnitude compared to that of the minicluster encounter, however, a large fraction of the sky observes effectively no radio flux.

\subsection{Time structure}
We now turn our attention to the short-term time structure of the radio signal generated from the axion clump-neutron star encounters (recall that the long-term time structure, depicted in \Fig{fig:long_time}, is dictated by the density profile of the object prior to in-fall, whereas the short-term time structure is dictated by the rotational period of the pulsar). For typical pulsars, this amounts to looking for variation in the flux density on timescales of $\sim 0.1 - 10$ seconds.

For the fiducial axion clump encounter, we generate 12 sky maps evenly spaced over the rotational period of the pulsar. We isolate a number of small regions on the sky $(\theta, \phi)$, and compute the flux density at that point in each sky map.

The top panel of \Fig{fig:timeD} illustrates the temporal behavior (as a function of rotational phase $\psi \equiv \Omega \times t$) of a typical minicluster-neutron star encounter, where we have taken $\theta_v = 0.3, 0.5,$ and 0.9 radians, and $\phi = 0$.  The solid lines are obtained by taking an azimuthal slice through the $t = 0$ sky map -- note that this is the temporal dependence that would arise if the phase space were isotropic (as \eg studied in~\cite{Witte:2021arp}); this approach continues to serve as a good approximation so long as the asymptotic phase space is approximately homogeneous, as is the case for a minicluster.  The horizontal error bars associated to each point in \Fig{fig:timeD} represent the width of the azimuthal bin over which the flux is averaged. The vertical bars, on the other hand, are a rough approximation of the statistical uncertainty (and are intended to illustrate the agreement between the two procedures); these are obtained by generating 6 sky maps at $t = 0$ (each with 5 million photons), taking 10 evenly spaced $\phi$ values at a fixed $\theta$, and determining the average standard deviation $\left< \sigma_S \right>$ of each sample (normalized to the mean flux). In reality, we plot $\sqrt{2} \times \left< \sigma_S \right> $ to account for the fact that this uncertainty enters both the discrete data points as well as the azimuthal slice (solid curve).

In the bottom panel of \Fig{fig:timeD} we show the temporal evolution of the axion star-neutron star encounter over a timescale slightly larger than the rotational period of the neutron star. In the case of the axion star, the inhomogeneity is so large that the azimuthal slice approximation is inapplicable, and thus we do not plot this quantity for comparison.  \Fig{fig:timeD} illustrates two important points: $(i)$ the signal remains strong throughout the encounter, but exhibits order-of-magnitude variations over the rotational period of the neutron star, and $(ii)$ similarly as for miniclusters, the peak signal strength depends on the viewing angle; however, since the axion star flux is more localised, this dependency is more pronounced.

\subsection{Line Width}

One of the defining features of these radio transients is the narrow spectral line, which is roughly centered about the axion mass, that arises as a result of the extremely low escape velocities of the axion clumps. The minimal value of the characteristic line width $f_\sigma$ is set by the typical velocity dispersion of axions prior to in-fall; it is well known, however, that the background plasma can induce significant spectral broadening after photon production, with the amplitude of the broadening depending on the rotational frequency and the characteristic scale of the conversion surface. Given that the signal to noise ratio scales with the $(f_\sigma)^{-1/2}$ (assuming an observing bandwidth $\lesssim f_\sigma$), designing an optimal radio search requires understanding the characteristic line width arising from these transient events.

\begin{figure}[t]
    \centering
    \includegraphics[width=0.48\textwidth]{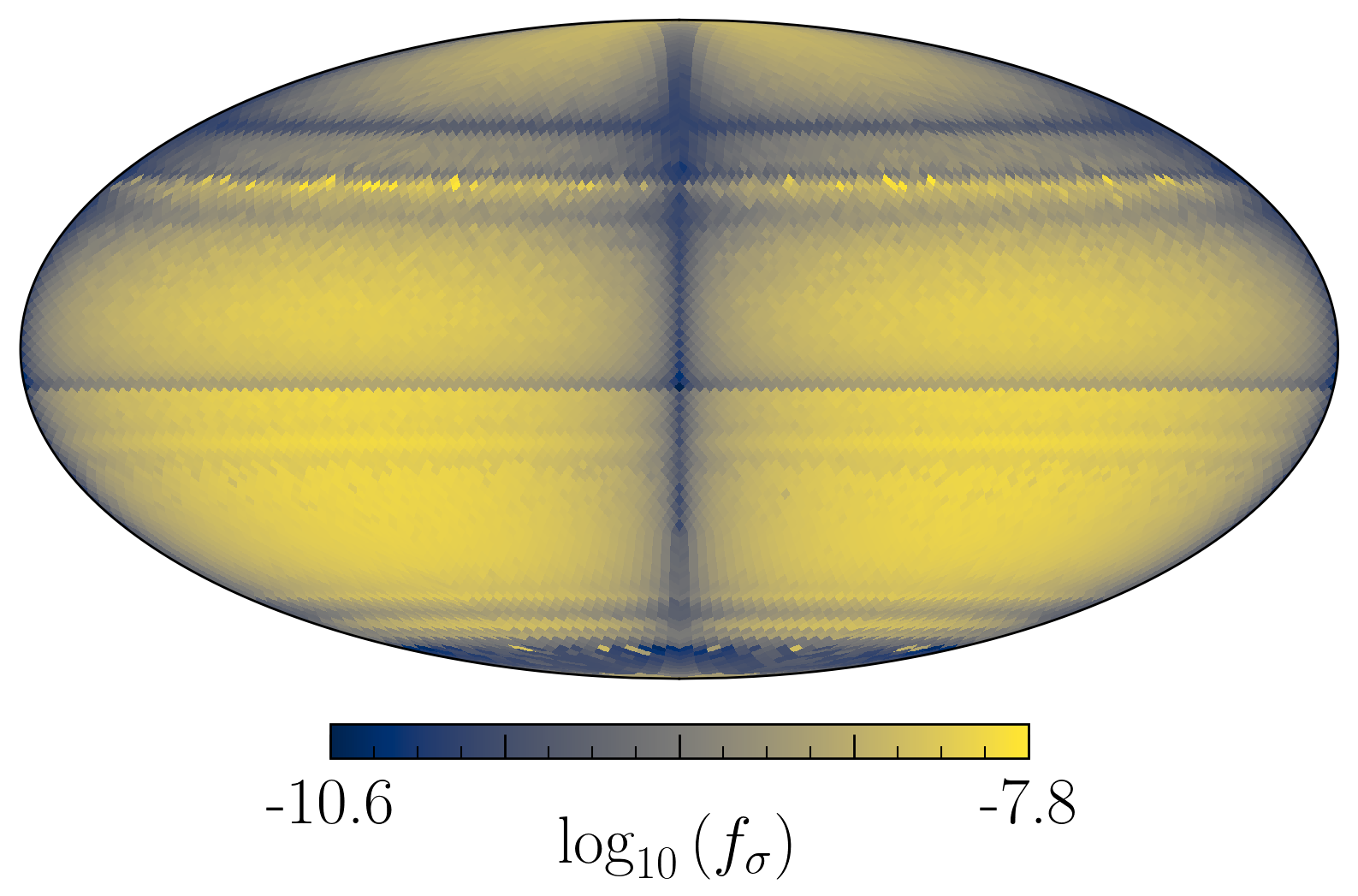}
    \caption{Same as top panel of \Fig{fig:skymap_width_amc_lc} (\ie an axion minicluster-neutron star encounter with relative angle between the minicluster velocity and the neutron star axis of rotation of $\theta_v = 0$), but taking the symmetry axis of the neutron star's magnetic field to be almost aligned with the neutron star's axis of rotation, $\theta_m = 0.001\,$rad. }
    \label{fig:ultra_narrow}
\end{figure}

In general, there are two relevant effects which can serve to broaden the line. The first is a shift in the central value of the line over the course of a rotational period (any observation averaging over long observation times would thus observe a broadened line). This effect, however, is in most cases strongly subdominant to the second effect, which is a pure broadening of the line (\ie the case in which the central value of the line $\bar{E}$ does not shift, but the width $f_\sigma$ grows). As a result, in what follows we neglect the former effect and define the net width to be determined by the standard deviation of the rate-weighted photon distribution in a given pixel (see \Eq{eq:fsig}). 

\begin{figure}[ht]
    \centering
    \includegraphics[width=0.49\textwidth]{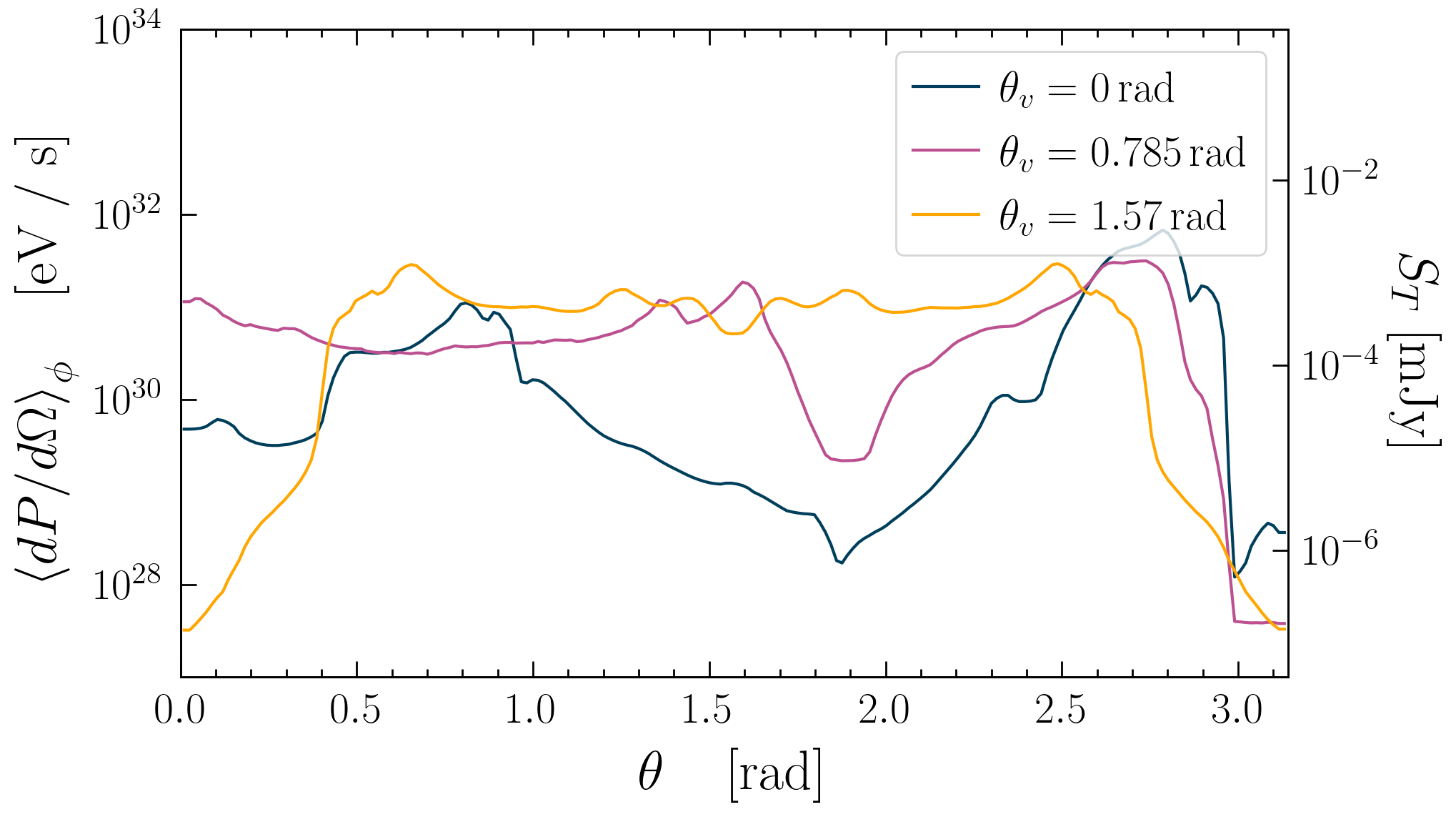}
    \caption{Radio flux (averaged over the azimuthal angle) $\left\langle dP/d\Omega\right\rangle_\phi$ as a function on viewing angle $\theta$ arising from a minicluster-neutron star encounter. The three lines illustrate the dependence on the direction of the relative minicluster-neutron star velocity $\theta_v$. }
    \label{fig:diff_p_v1}
\end{figure}

The results for the minicluster and axion star encounters are shown in \Fig{fig:skymap_width_amc_lc} and \Fig{fig:skymap_width_as}, respectively. As before, sky maps are shown for three different values of the encounter angle $\theta_v$. The maximum width across all maps tends to be $\sim \mathcal{O}(10^{-6} \, m_a)$, however some pixels produce lines that are orders of magnitude narrower than this value.

\begin{figure*}
    \centering
    \includegraphics[width=0.48\textwidth]{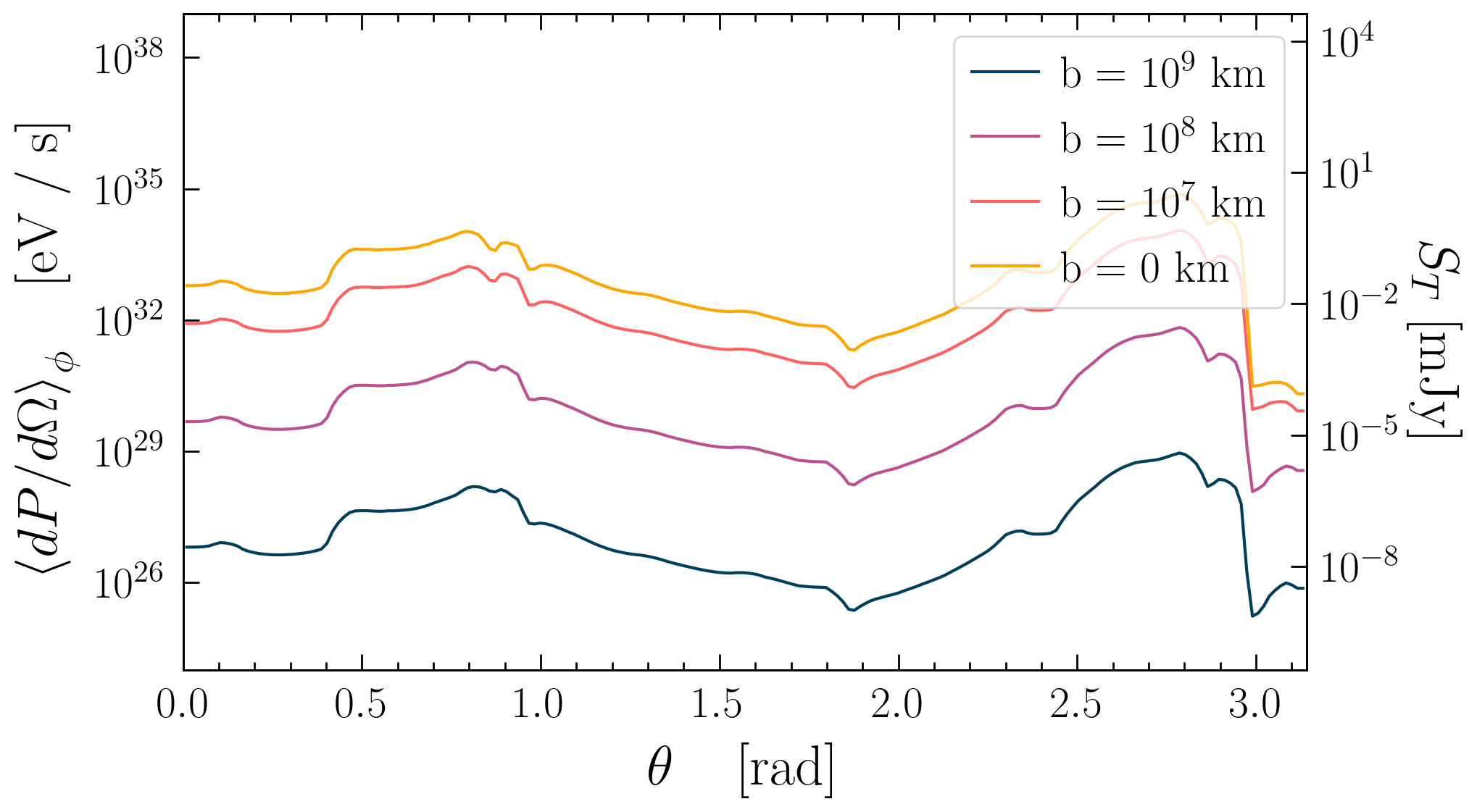}
    \includegraphics[width=0.48\textwidth]{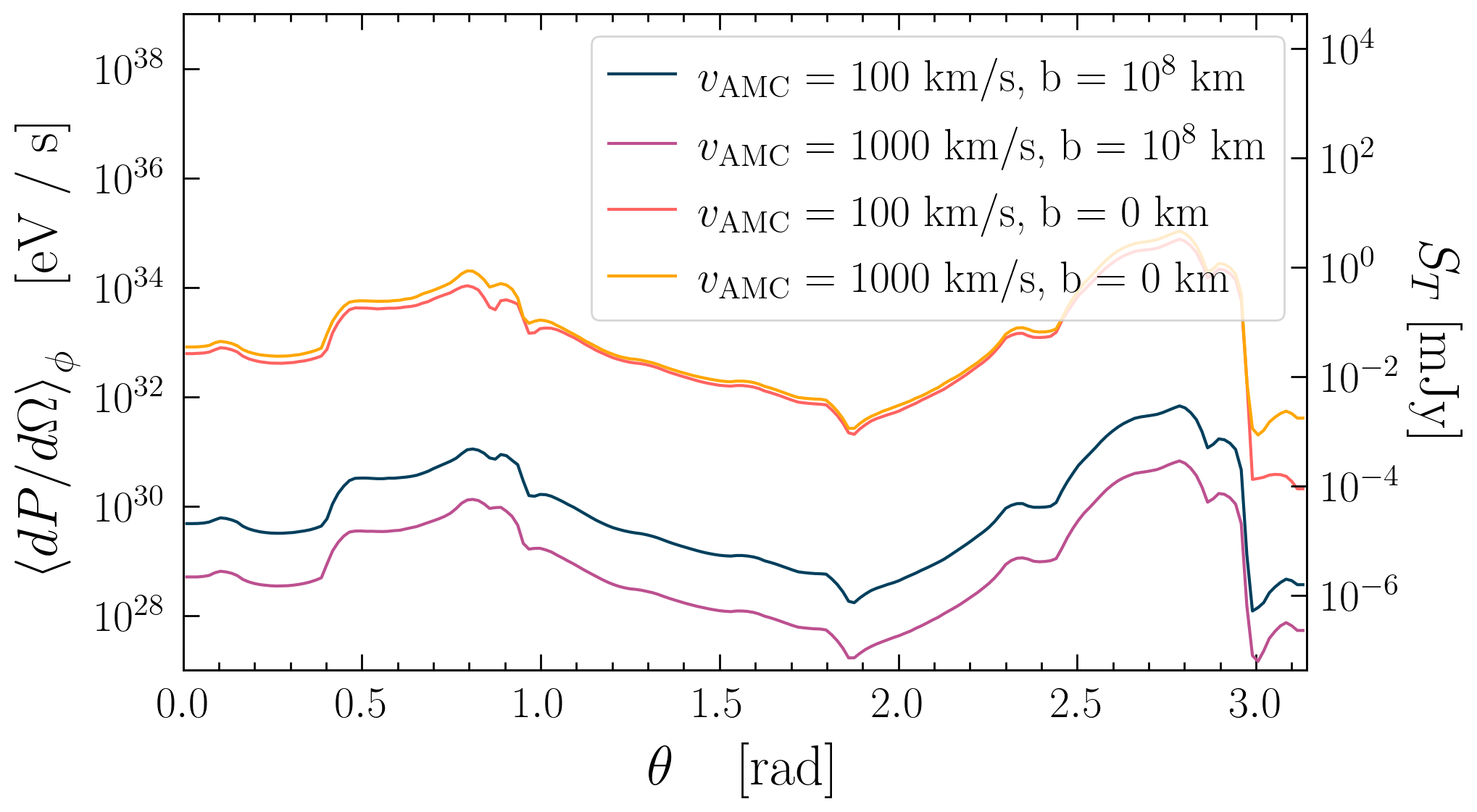}
    \caption{Same as \Fig{fig:diff_p_v1}, but illustrating the dependence on the impact parameter ($b$) and the magnitude of the relative minicluster-neutron star velocity ($v_{\rm AMC}$). }
    \label{fig:diff_p_v2}
\end{figure*}

It is worth highlighting here a major difference between the scenario of axion clump-neutron star transient encounters and the case in which radio lines are sourced from a smooth background distribution of axions. In the case of the latter, the minimum line width is set by the typical energy in the rest frame of the halo $\sim m_a v_0^2 / 2$, where $v_0 \sim 10^{-3}$. Here, the minimum line width is set by the velocity dispersion of the axion clump, which can be many orders of magnitude smaller. As a result, transient encounters with highly aligned or slow rotators (where plasma broadening effects are heavily reduced) can produce far narrower, and more distinctive, lines. As an example, we plot in \Fig{fig:ultra_narrow} the characteristic width that would arise from a nearly aligned rotator, with $\theta_m = 0.001$ radians, showing that typical values are on the order of $\sim 10^{-8} m_a$.

\subsection{Sensitivity to Impact Parameter and Relative Velocity}

Thus far, we have kept the impact parameter and the magnitude of the relative axion clump-neutron star velocity fixed. Here, we illustrate the sensitivity of the flux density to reasonable variations in these parameters.

In the case of the minicluster-neutron star encounter, we choose to plot the phase-averaged differential power $\left< dP/d\Omega \right>_\phi \equiv \frac{\omega}{2\pi} \int d\phi \,  \frac{dP}{d\Omega}(\phi, \theta)$ as a function of viewing angle $\theta$ rather than the sky map, as this quantity is closer to being directly analogous to what an observer would measure (note that while this is not exactly equivalent to the period-averaged flux, \Fig{fig:timeD} verifies that is a very good approximation for the typical minicluster). As an illustrative example, we plot $\left< dP/d\Omega \right>_\phi$ in \Fig{fig:diff_p_v1} (as well as the corresponding flux density, for the fiducial observation parameters) for the three minicluster-neutron star encounters discussed in the sections above, such that a direct comparison can be made with \eg the sky maps of \Fig{fig:skymap_amc_lc}.

We illustrate in \Fig{fig:diff_p_v2} the dependence on the impact parameter (left), and on both the impact parameter and the relative minicluster-neutron star speed (right). To a large degree, variations in the impact parameter amount to a net overall scaling of the observed flux density. The relative minicluster speed has a minimal impact for head-on collisions, while at large impact parameters (which are far more probable), slower encounter speeds lead to larger flux densities (a natural expectation of Louisville's theorem). Importantly, both the impact parameter and the relative velocity can strongly impact the long-term time evolution of the transient event shown in \Fig{fig:long_time}, significantly altering both the total length of the transient signal as well as the relative time evolution.

\begin{figure*}
    \centering
    \includegraphics[width=0.45\textwidth]{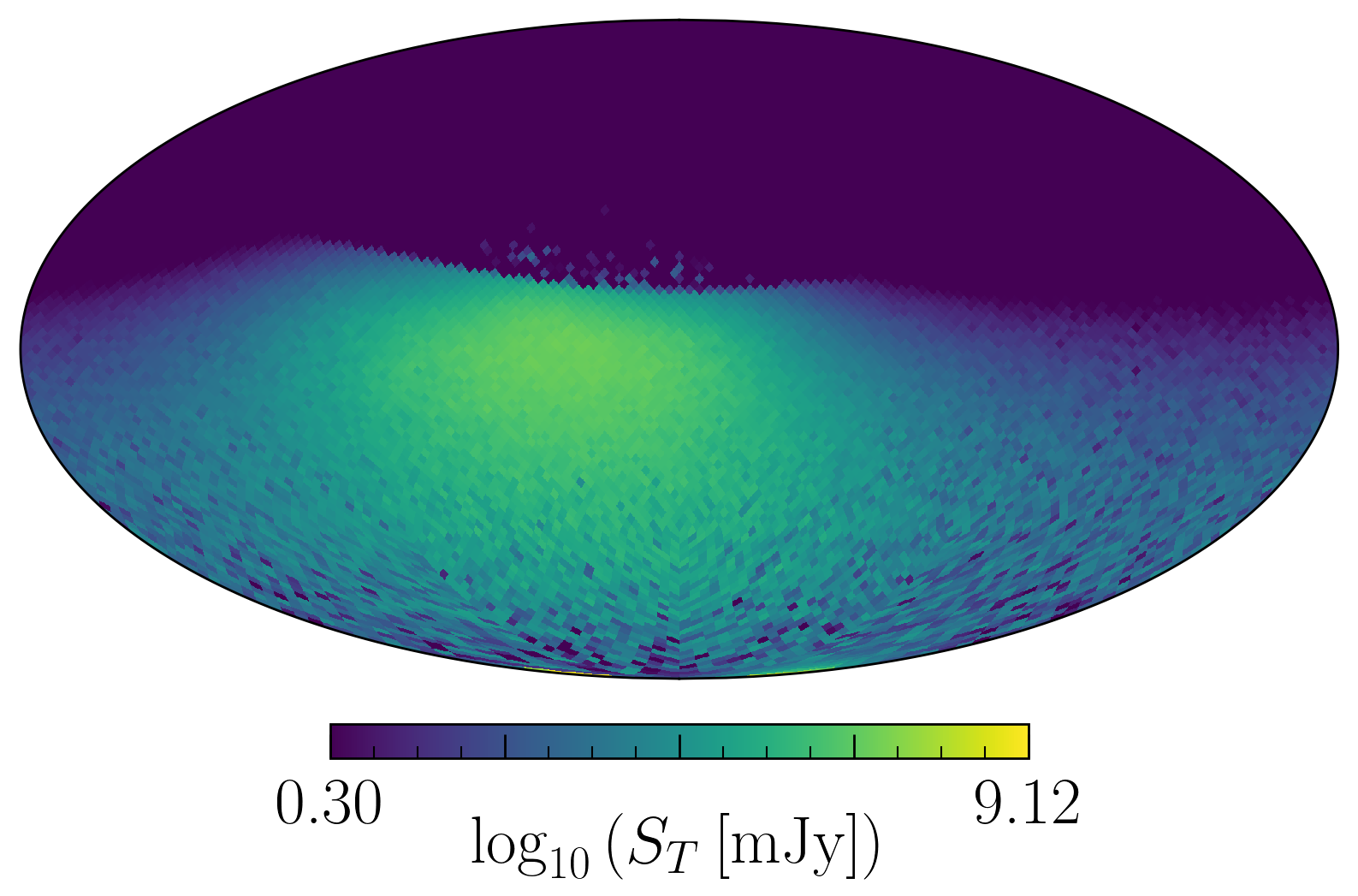}
    \includegraphics[width=0.45\textwidth]{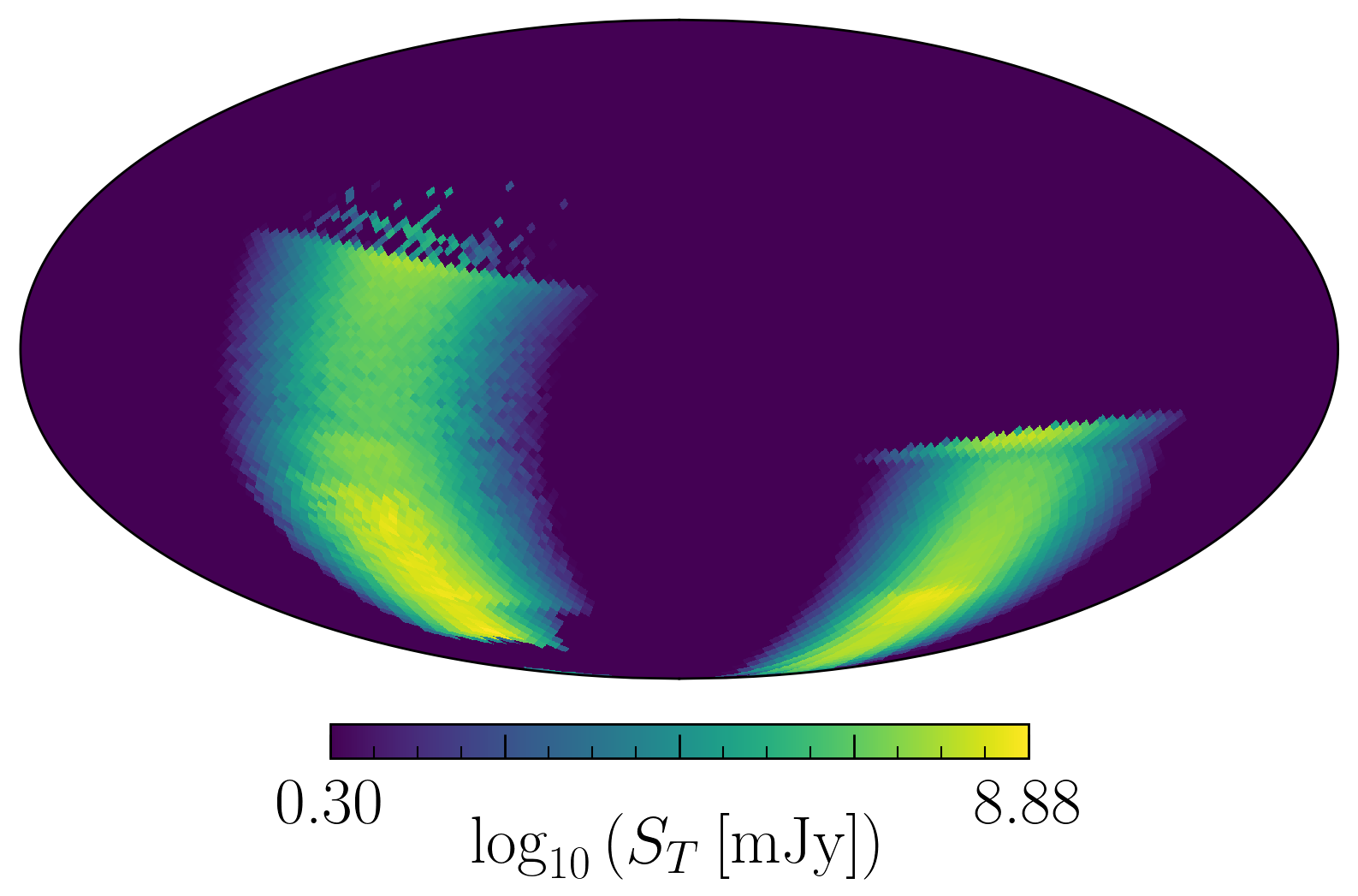}    
    \caption{Sky maps of the radio flux $S_T$ arising from axion star-neutron star encounters, varying the impact parameter $\vec{b} = \vec{0}$ (left) and $\vec{b} = 5 \times 10^3 \, \hat{y}\,$km (right).}
    \label{fig:AS_varyb}
\end{figure*}

Axion star-neutron star encounters at non-vanishing impact parameter exhibit a subtle dependence on the internal structure of the axion star, as we now discuss. A (disrupted) axion star only reaches the conversion surface if the impact parameter is sufficiently small: with a speed of $10^3$ km/s at the Roche radius, the relevant impact parameters are $|\vec{b}| \lesssim R_{\rm AS}$, i.e.~all encounters that generate a signal are nearly head-on. For axion star-neutron star encounters, one cannot approximate the period-averaged flux using $\left< dP/d\Omega \right>_\phi$, so in \Fig{fig:AS_varyb}, we show sky maps illustrating the impact of shifting the impact parameter on an axion star-neutron star encounter, with the left map showing a head-on collision and the right a skirting trajectory (i.e.~a collision which nearly misses). The anisotropy in the case of the head-on collision is somewhat similar to the fiducial model shown in \Fig{fig:skymap_as}, producing a comparable flux density but in a slightly more homogeneous manner. The skirting collision, on the other hand, is notably stronger at maximum, but is more anisotropic, producing a flux over only a small fraction of the sky. That off-set trajectories can create a stronger signal than head-on collisions may be surprising, given that the mean turn radius (point of closest approach) increases monotonically with the impact parameter. To explain this result, we note that the intrinsic velocity distribution within the axion star gives a preference for a non-vanishing speed in the $\hat x$ and $\hat y$ directions. This is sufficient to cause axions with small impact parameters at the Roche radius to preferentially miss the resonant surface, but modestly offset axions can be more favorably oriented to pass close to the neutron star. The strength of this effect is sensitive to the detailed (and unknown) intrinsic velocity distribution. Similar behavior can be seen in the scaling of of the temporal evolution of the axion star neutron star encounter shown in \Fig{fig:long_time}.

\begin{figure*}
    \centering
    \includegraphics[width=0.45\textwidth]{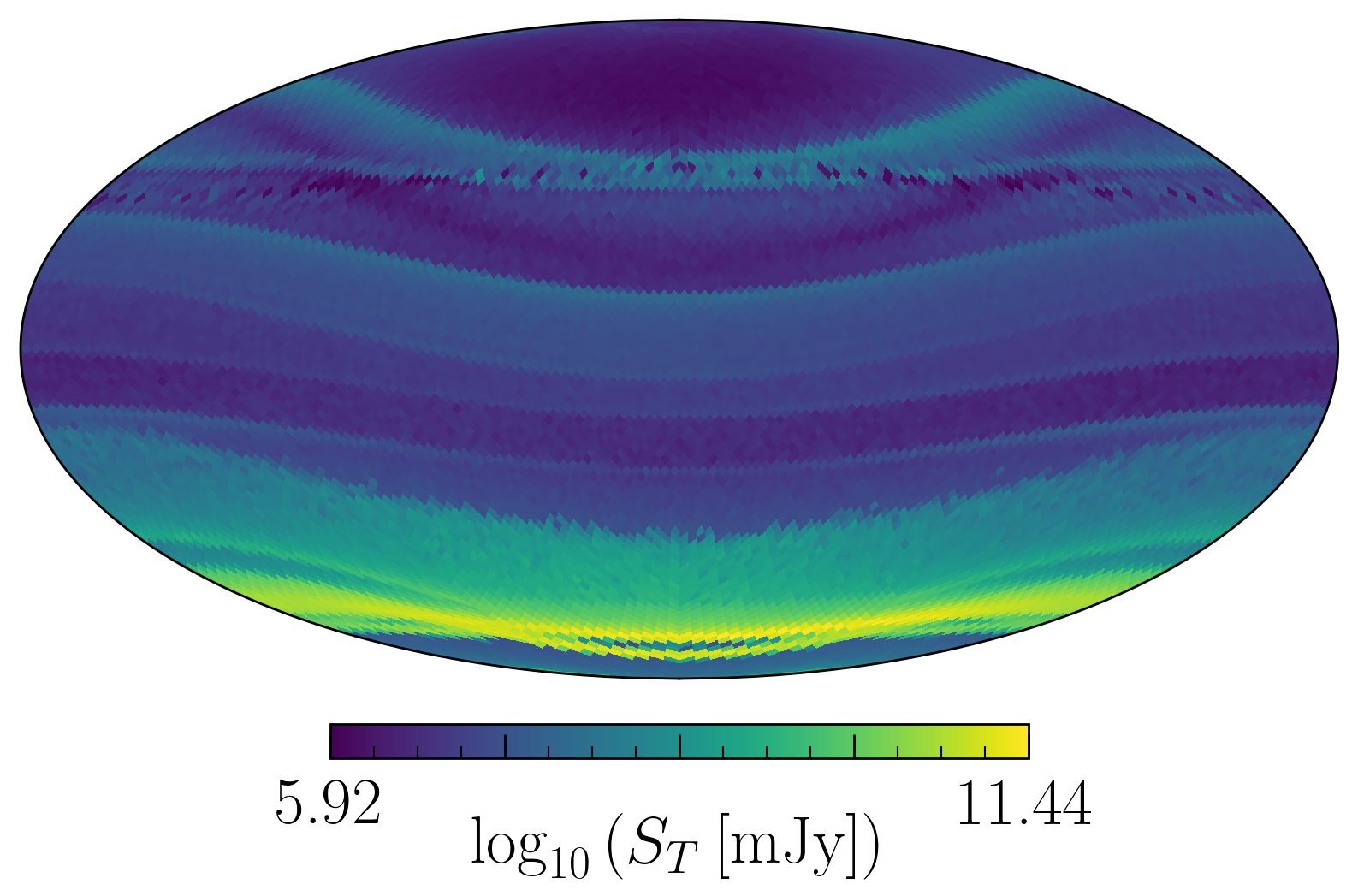}
    \includegraphics[width=0.45\textwidth]{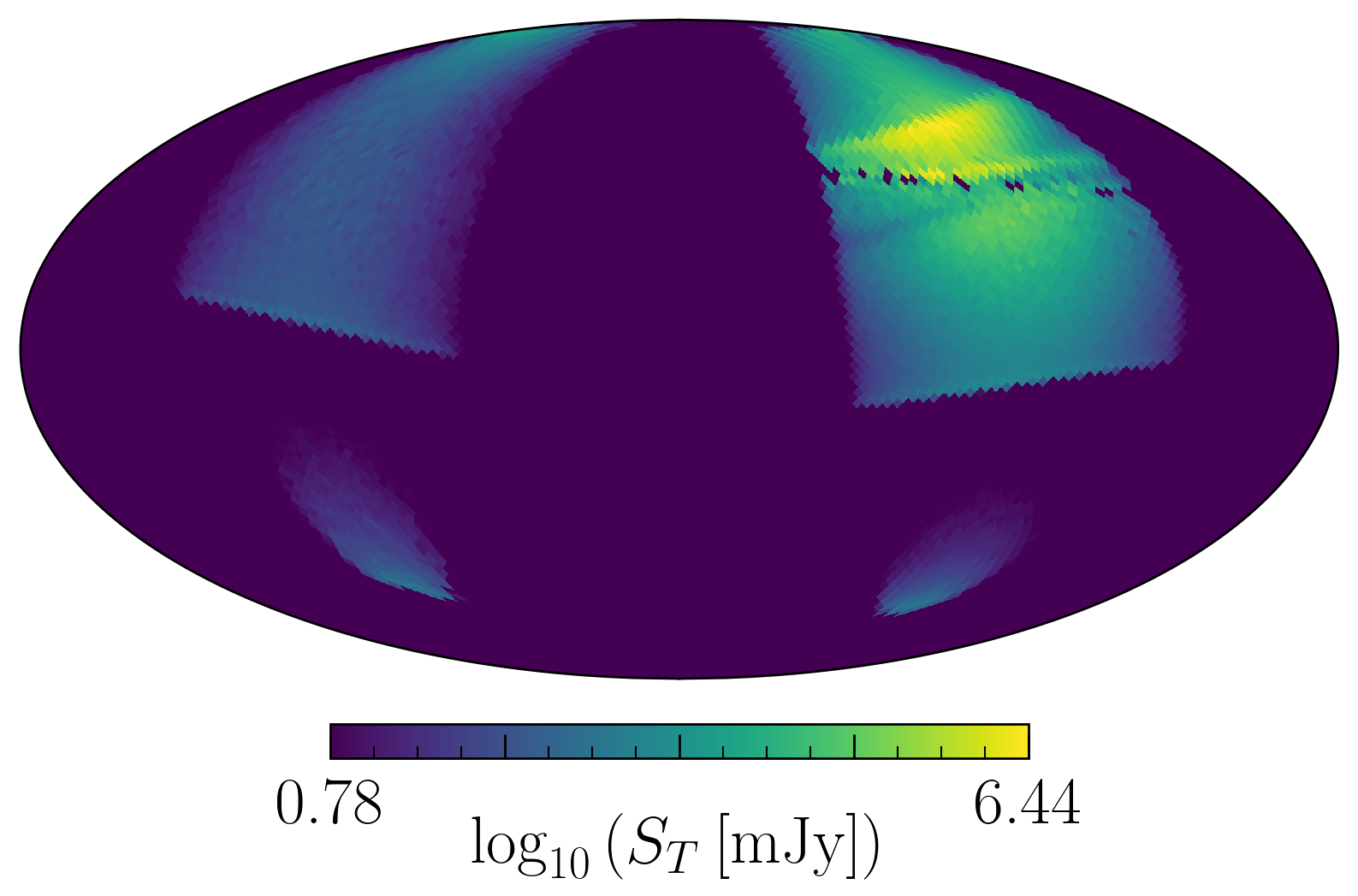}
    \caption{Sky maps of the radio flux $S_T$ arising from axion star-neutron star encounters, taking $|v_{\rm AS}| = 10^3\,$km/s and varying the impact parameter $\vec{b} = \vec{0}$ (left) and $\vec{b} = 5 \times 10^3 \, \hat{y}\,$km (right). }
    \label{fig:AS_varyV}
\end{figure*}

Finally, \Fig{fig:AS_varyV} shows the effect of varying the relative speed and impact parameter of an axion star-neutron star encounter, with the speed taken to be $10^3\,$km/s instead of our fiducial assumption of 100\,km/s -- as before, the impact parameters are taken to be $0$ (left) and $5 \times 10^3 \, \hat{y}\,$km (right). The left panel shows regions of extremely high flux (exceeding at some point $10^{11}\,$mJy), and increased sky coverage when compared with fiducial sky map of \Fig{fig:skymap_as}. 

\section{Conclusions} \label{sec:conc}

In this work we have investigated the  signatures arising from the disruption and stripping of axion miniclusters and axion stars by neutron stars; these rare encounters are capable to producing bright transient radio lines, with frequencies roughly centered about the axion mass, that endure for timescales spanning from seconds to months. Until now,  there has not been a complete description of the gravitational in-fall, photon production, and photon propagation, making it difficult to estimate the strength of the signal, the spectral characteristics, beaming effects, etc. Here, we generate the first end-to-end pipeline to follow the evolution of this system, starting from the axion phase space in the asymptotic past and going to the final photon distribution in the asymptotic future. This allows us to answer for the first time: what does the radio signal from a transient axion clump-neutron star encounter actually look like?

For minicluster-neutron star encounters, we find the anisotropy of the radio flux across the sky to be significant, with our fiducial model showing variations of the flux across the sky by 4-5 orders of magnitude. Not surprisingly, for the case of axion star-neutron star encounters the anisotropy is even more pronounced, with large fractions of the sky receiving no observable flux. For rotational frequencies and misalignment angles typical of active pulsars, we find that the line width is dominated by plasma broadening effects (thus, the relative line width of the signal ranges from $f_\sigma \sim 10^{-6}\textit{--}10^{-4}$); however, for aligned and slow rotators, the width of the line can many orders of magnitude less, introducing the possibility that dead neutron stars may produce more distinctive signatures.

We also characterize the long-term and short-term time evolution of both minicluster and axion star  encounters, showing that miniclusters can naturally produce time variation at least at the order of magnitude level over a rotational period, while the flux from axion stars can vary by orders of magnitude on sub-second timescales. Furthermore, we show that the strength, anisotropy, and time dependence of encounters can be strongly dependent on the relative velocity and the impact parameter of the encounter itself.

In this work we do not attempt to make any statement on the transient signals arising from the broader population of axion clump-neutron star encounters; rather, we focus on developing the formalism and tools needed to characterize the properties from a particular encounter. In addition to the methods developed here, assessing the observability (and optimizing observing strategies) of axion clumps requires a more detailed understanding of the properties and distributions of the minicluster, axion star, and neutron star populations; owing to the added complexity, we reserve such a study for future work.

Nevertheless, the results shown here have demonstrated that there exists a strong sensitivity of the radio flux to the encounter parameters, which likely necessitates large encounter rates, as strong but rare events are likely to give better sensitivity to the axion-photon coupling. In future work we intend to look at the projected transient event rate arising in nearby galaxies; this will require a detailed treatment of neutron star population synthesis and an understanding of the tidal stripping and disruption of miniclusters arising from stellar encounters. Collectively, these studies will form the basis which will allow us to address the greater challenge of understanding the observability of axion clump-neutron star transient events, with a particular focus on how to develop optimized search strategies to be employed in future radio observations.

\acknowledgments{}
The authors would like to thank Christoph Weniger for his participation in the initial stage of this work.
SJW is supported by the European Research Council (ERC) under the European Union's Horizon 2020 research and innovation programme (Grant agreement No. 864035 -- Undark) and the Netherlands eScience Center, grant number ETEC.2019.018. 
The work of SB is supported by NSF grant No. PHYS-2014215, DoE HEP QuantISED award No. 100495, and the Gordon and Betty Moore Foundation Grant No. GBMF7946. AJM is supported by the European Research Council under Grant No. 742104 and by the Swedish Research Council (VR) under Dnr 2019-02337 “Detecting Axion Dark Matter In The Sky And In The Lab (AxionDM)”. Fermilab is operated by Fermi Research Alliance, LLC under Contract No. DE-AC02-07CH11359 with the United States Department of Energy.
The work of DM and GS is supported by the European Research Council under Grant No. 742104 and by the Swedish Research Council (VR) under grants 2018-03641 and 2019-02337.

\bibliography{biblio}

\end{document}